\documentclass[11pt,a4paper]{article}
\usepackage{jheppub}
\usepackage{amsmath,amssymb,comment,color,braket,epsfig,graphicx,hyperref,here,natbib,slashed,subfigure,mathrsfs}
\graphicspath{{./Figures/}}
\def\nn{\nonumber}
\def\D{\mathrm{d}}
\def\beq{\begin{equation}}
\def\eeq{\end{equation}}
\def\bea{\begin{eqnarray}}
\def\eea{\end{eqnarray}}
\begin{document}
\title{Quark-hadron duality for heavy meson mixings\\
in the 't Hooft model}
\author{Hiroyuki Umeeda}
\emailAdd{umeeda@gate.sinica.edu.tw}
\affiliation{Institute of Physics, Academia Sinica, Taipei 11529, Taiwan, Republic of China}
\date{\today}
\abstract{We study local quark-hadron duality and its violation for the $D^0-\bar{D}^0$, $B^0_d-\bar{B}^0_d$ and $B^0_s-\bar{B}^0_s$ mixings in the 't Hooft model, offering a laboratory to test QCD in two-dimensional spacetime together with the large-$N_c$ limit. With the 't Hooft equation being numerically solved, the width difference is calculated as an exclusive sum over two-body decays. The obtained rate is compared to inclusive one that arises from four-quark operators to check the validity of the heavy quark expansion (HQE). In view of the observation in four-dimensions that the HQE prediction for the width difference in the $D^0-\bar{D}^0$ mixing is four orders of magnitude smaller than the experimental data, in this work we investigate duality violation in the presence of the GIM mechanism. We show that the order of magnitude of the observable in the $D^0-\bar{D}^0$ mixing is enhanced in the exclusive analysis relative to the inclusive counterpart, when the 4D-like phase space function is used for the inclusive analysis. By contrast, it is shown that for the $B^0_d-\bar{B}^0_d$ and $B^0_s-\bar{B}^0_s$ mixings, small yet non-negligible corrections to the inclusive result emerge, which are still consistent with what is currently indicated in four-dimensions.}
\keywords{Heavy Quark Physics, 1/N Expansion, Nonperturbative Effects}
\maketitle
\section{Introduction}\noindent
The theory of heavy quark physics, established since 1980s, has already experienced its mature stage. While its early development is characterized particularly by the heavy quark symmetry, nowadays it is turned into a systematic way to handle non-perturbative aspects of quantum chromodynamics (QCD). Equipped with Wilson's operator product expansion (OPE) \cite{Wilson:1969zs,Wilson:Proc,Wilson:1973jj} (the ideas were adopted to QCD in Refs.~\cite{Shifman:1978bx, Shifman:1978by, Novikov:1984rf}), certain processes in the deep Euclidean domain are factorized into short and long distance objects. The former is calculated via perturbation theory while the latter is evaluated by non-perturbative methods such as lattice QCD. The OPE formula is then converted into one in the Minkowskian domain, on which physical processes of interest lie, via the analytic continuation. As a result, the observables are expanded by the inverse of heavy quark mass, $1/m_Q$. This methodology, referred to as the heavy quark expansion (HQE) \cite{Bigi:1992su, Bigi:1992ne, Blok:1992hw, Blok:1992he} (see, \textit{e.g.,} Refs.~\cite{Bigi:1997fj, Lenz:2014jha} for reviews), is quite successful in describing inclusive processes for $b$ quark. The current results for the lifetime ratios of $b$-hadrons \cite{Lenz:2014jha, Kirk:2017juj,Cheng:2018rkz} and the width difference in the $B^0_{s}-\bar{B}^0_{s}$ mixing \cite{Lenz:2019lvd} show an excellent agreement with the Heavy Flavor Averaging Group (HFLAV) data \cite{Amhis:2019ckw}.\par
In contrast to the successful aspects of HQE for $b$ quark, there exists two-fold complexity for treating $c$ quark: (1) charm might be possibly too light for applying HQE and (2) due to Glashow-Iliopoulous-Miani (GIM) mechanism \cite{Glashow:1970gm}, observables undergo severe cancellation unlike the milder one for $b$ quark. Due to the latter, specifically relevant for flavor-changing neutral current (FCNC) processes, observables are subject to the suppressions of SU(3) breaking \cite{Kingsley:1975fe} and/or the tiny product of Cabibbo-Kobayashi-Maskawa (CKM) matrix \cite{Cabibbo:1963yz, Kobayashi:1973fv} elements, $V_{cb}^*V_{ub}$.\par
One of such notoriously difficult FCNC processes of $c$ quark is the $D^0-\bar{D^0}$ mixing\footnote{For the experimental side, the first evidence was found by Belle \cite{Staric:2007dt} and BABAR \cite{Aubert:2007wf} collaborations in 2007. Subsequent confirmation was made by CDF \cite{Aaltonen:2007ac} and LHCb \cite{Aaij:2013wda} experiments. Currently, the average over large datasets \cite{Amhis:2019ckw} show that the zero values of the mixing parameters are excluded by more than $11.5\sigma$ \cite{Amhis:2019ckw}, so that the occurrence of the $D^0-\bar{D^0}$ mixing has been firmly verified. See Refs.~\cite{Amhis:2019ckw, Lenz:2020awd} for the detail of the experimental status and references therein.}, that proceeds via $\Delta C=2$ transition (see Refs.~\cite{Burdman:2003rs, Lenz:2020awd} for reviews). Two possible methods to calculate the $D^0-\bar{D^0}$ mixing exist in the literature: exclusive and inclusive approaches, where the latter is based on HQE. In the exclusive approach \cite{Falk:2001hx,Wolfenstein:1985ft,Donoghue:1985hh,Colangelo:1990hj,Buccella:1994nf,Kaeding:1995zx,Falk:2004wg,Cheng:2010rv,Gronau:2012kq,Jiang:2017zwr}, the experimental data of hadronic decays are utilized so that the relevant long-distance effect can be properly extracted. The modern analyses \cite{Cheng:2010rv,Jiang:2017zwr} showed that two-body decays of $D^0$ meson accommodate roughly a half of the width difference although there lies a difficulty in handling other multi-body modes. Hence, while the order of magnitude of the width difference was reproduced, the quantitative agreement is still not realized in the exclusive approach.
\par
On the other hand, the situation of the inclusive approach to the $D^0-\bar{D^0}$ mixing is somewhat different from that to the exclusive one. Owing to the severe GIM cancellation, the inclusive values of the mass and width differences are considerably suppressed, as can be seen from formulae obtained by the box diagrams in Refs.~\cite{Hagelin:1981zk, Cheng:1982hq, Buras:1984pq, Datta:1984jx} and also by the heavy quark effective field theory in Refs.~\cite{Georgi:1992as,Ohl:1992sr}. The later update including next-to-leading order (NLO) corrections, obtainable from proper replacement in the $B^0-\bar{B^0}$ mixing \cite{Beneke:1996gn, Beneke:1998sy,Dighe:2001gc,Ciuchini:2003ww} (see also \cite{Petrov:1997ch}), gives the width difference about four orders of magnitude smaller \cite{Golowich:2005pt, Bobrowski:2010xg} than the HFLAV data \cite{Amhis:2019ckw}. This huge discrepancy is to be contrasted with the exclusive approach, in which the order of magnitude is accommodated. Another point to be mentioned is that the HQE prediction for $\tau(D^+)/\tau(D^0)$ in Ref.~\cite{Kirk:2017juj} is in agreement with the HFLAV data \cite{Amhis:2019ckw}, albeit the huge uncertainty in the theoretical side, indicating that the HQE for $c$ quark is more or less meaningful in the processes without GIM cancellation.
\par
In order to interpret the aforementioned disagreement, several possibilities are discussed in the literature\footnote{See the status summarized in Ref.~\cite{Jubb:2016mvq}.}: first one is attributed to the contributions of higher dimensional operators, potentially leading to an enhancement, as discussed in \cite{Georgi:1992as, Bobrowski:2010xg, Bigi:2000wn, Falk:2001hx}. For further clarifying this possibility, one should calculate a number of non-perturbative matrix elements for $D=9, 12$ operators. Indeed, a new physics contribution is considered a candidate for explaining the gap. See, \textit{e.g.}, Refs.~\cite{Golowich:2006gq,Golowich:2007ka, Golowich:2009ii, Gedalia:2009kh} for the studies in the context of new physics. A subtle point discussed in the recent work \cite{Lenz:2020efu} is that if one adopts $\mu_1$, a scale at which the bi-local process induced by the $\Delta C=1$ oparators is calculated, different for individual internal quark contributions, the sufficient enhancement is realized after taking sum over flavors. In this respect, a natural question might be how large the next-to-next-to-leading order (NNLO) QCD corrections \cite{Asatrian:2017qaz, Asatrian:2020zxa} will be after its completion. Furthermore, another recent study \cite{Li:2020xrz} where the dispersion relation is regarded as a constraining equation to determine the width difference at low energies indicated that the inclusive approach potentially leads to an enhancement.
\par
An alternative possibility to interpret the discrepancy is violation of quark-hadron duality.\footnote{In the past, duality violation was considered crucial to explain the lifetime ratio of $\tau(\Lambda_b)/\tau(B_d)$ although it was falsified due to the update in experimental data.} The notion of duality is originated from investigations due to Bloom-Gilman \cite{Bloom:1970xb, Bloom:1971ye} and Poggio-Quinn-Weinberg \cite{Poggio:1975af} stating that inclusive hadronic cross sections at high energies are described by the quark-gluon picture. The case with smearing observables over energies is referred to as ``global duality,'' while one without smearing is called ``local duality.'' The difficulty in handling duality violation is traced back to the truncations of perturbative series for $\alpha_s$ and OPE. Specifically, the proliferation of Feynman diagrams gives rise to factorial divergence, which is not included in the practical version of OPE. In addition, it is known that renormalons \cite{Beneke:1998ui}, referring to countributions from particular diagrams, also lead to the factorial divergence. Furthermore, the series from OPE is divergent \cite{Shifman:1994yf, Shifman:1995mt} as well. Due to those corrections, the higher order perturbative series should be truncated at an optimal order, leaving an uncertainty in the perturbative prediction. Thus, the accuracy of the resultant HQE, intrinsically replying on the truncated series with the analytic continuation, is limited up to those non-perturbative effects. See, \textit{e.g.,} Refs.~\cite{Shifman:2000jv, Bigi:2001ys} for further details regarding duality violation.
\par
While obviously a first principle method in the Minkowskian domain is preferable, duality violation is hard to quantify as long as one depends on the truncated perturbative series (for wording of ``duality violation,'' we follow the clear-cut definition due to Shifman \cite{Shifman:2000jv}, referring to the error beyond \textit{the natural uncertainties} of truncated series from $\alpha_s$ and OPE). In the literature, certain dynamical mechanisms are considered as models of duality violation. These approaches are: (a) instanton-based model in Refs.~\cite{Chay:1994si,Chay:1994dk,Falk:1995yc,Chibisov:1996wf} and (b) resonance-based model in Refs.~\cite{Shifman:1994yf, Shifman:1995mt,Zhitnitsky:1995qa, Blok:1997hs, Colangelo:1997ni, Grinstein:1997xk, Bigi:1998kc, Grinstein:1998gc, Bigi:1999fi, Bigi:1999qe, Burkardt:2000uu,Burkardt:2000ez,Lebed:2000gm,Beane:2001uj,Grinstein:2001zq,Grinstein:2001nu,Mondejar:2006ct,Mondejar:2008pi,Mondejar:2009td} and also in Ref~\cite{Golowich:1998pz}.\footnote{Another pure phenomenological approach based on the simple model \cite{Jubb:2016mvq} showed that $20\%$ violation of duality can account the width difference of the $D^0-\bar{D^0}$ mixing. See also Refs.~\cite{Gambino:2020crt, Fukaya:2020wpp} for the recent works in lattice QCD to calculate inclusive processes.}
For (a), the usual perturbative analysis is replaced by one in the medium of (fixed-sized) instanton, classical solution to Yang-Mills equations in Euclidean space \cite{Belavin:1975fg}. This procedure leads to the contribution of {\it finite distance singularity} from the quark Green function, in addition to the practical OPE as the short-distance expansion, and gives a possible duality violating term that has an exponential-like function form. By performing analytic continuation to the Minkowskian domain, an oscillatory correction to the practical OPE arises when quark mass is not heavy enough.
\par
As for (b), duality violation is studied on the basis of the tower of hadronic excited states that follow the linear Regge trajectory and the large-$N_c$ limit (the finite correction from $1/N_c$ can be also included). This was considered for the hadronic vacuum polarization in Ref.~\cite{Blok:1997hs}. By summing over each hadronic propagator, one finds that the vacuum polarization is recast into Euler's $\psi$ function, whose asymptotic expansion leads to {\it the OPE series}. By comparing the hadronic result and the OPE series, where the latter is truncated in practice, one can investigate duality although for the vacuum polarization, either smearing or $1/N_c$ correction should be taken into account to gain a reasonable result, since local duality is maximally violated even for large energies for this case.
\par
Resonance-based investigation of duality is greatly facilitated with the help of the 't Hooft model \cite{tHooft:1973},  $1+1$ dimensional SU($N_c$) gauge theory in the large-$N_c$ limit \cite{tHooft:1973alw,Coleman:1985,Manohar:1998xv,tHooft:2002ufq}, in which case only the planar diagrams give non-vanishing contributions. The Bethe-Salpeter equation \cite{Nambu:1997vt, Salpeter:1951sz} in the light-cone gauge leads to a relation constraining wave functions and masses of mesons, the so-called 't Hooft equation. Being solvable, the equation unambiguously determines the properties of mesons in this formalism, thereby offering a useful laboratory to examine the non-perturbative dynamics of strong interaction. The (asymptotic) linear Regge trajectory, a key ingredient in (b), can be demonstrated in the model. Supported by such tractable features, discreteness of the mass spectra is shown mathematically \cite{Federbush:1976eh}, as is required by confinement. Posterior to the original work \cite{tHooft:1973}, the scattering amplitude, discussion in the axial gauge, chiral symmetry breaking, simulation on the lattice (with finite $N_c$), generalized parton distribution functions, weak decays of heavy quark, \textit{etc.}, are investigated in Refs.~\cite{Callan:1975ps, Einhorn:1976uz, Pak:1976dk,Hanson:1976ey, Bars:1977ud, Brower:1978wm,Zhitnitsky:1985um, Li:1986gf, Li:1987hx, Huang:1988br,Burkardt:1992qm,Burkardt:1991ea, Jaffe:1991ib,Grinstein:1992ub, Grinstein:1994nx,Barbon:1994au,Aoki:1995dh,Krauth:1996dg,Abdalla:1998sg, Abdalla:1999av,Armoni:2000uw,Berruto:2002gn,Grinstein:2006pz,
Mondejar:2008dt,Grinstein:2008wm,Glozman:2012ev, Jia:2017uul,Jia:2018qee}. Particularly noteworthy is that the intermediate meson pole contribution to the heavy-to-light form factor is demonstrated for any heavy quark mass, and the correction to the approximation is also determined, so that QCD dynamics in heavy quark decays can be clarified in $1+1$ dimensions \cite{Grinstein:1994nx}. Numerical \cite{tHooft:1973, Hanson:1976ey, Brower:1978wm, Huang:1988br, Jaffe:1991ib, Krauth:1996dg,Armoni:2000uw, Fonseca:2006au}, semi-analytical \cite{Harada:1997kq} and analytical \cite{Lewy, Hildebrandt1,Hildebrandt2,Hildebrandt3,Bruning, Fateev:2009jf, Ziyatdinov:2010vg, Zubov:2015ura} methods to obtain solutions to the 't Hooft equation are investigated in the vast literature.
\par
The mentioned tractable features of the 't Hooft model enable us to test quark-hadron duality. In the previous studies, this test is applied for hadronic spectral density functions \cite{Zhitnitsky:1995qa, Blok:1997hs, Bigi:1998kc, Lebed:2000gm} related to $e^+e^-$ annihilation and $\tau$ decays, deep inelastic scattering \cite{Mondejar:2008pi,Mondejar:2009td} and heavy meson decays \cite{Zhitnitsky:1995qa, Grinstein:1997xk, Bigi:1998kc, Grinstein:1998gc, Bigi:1999fi, Bigi:1999qe,Burkardt:2000ez,Lebed:2000gm,Grinstein:2001zq,Mondejar:2006ct}. Some of those references analytically gave the oscillating behavior for process rates, which is not captured in the practical OPE, as the energy/heavy quark mass is lowered. Thus, it is broadly considered that the 't Hooft model offers one certain methodology to reliably analyze duality violation while how the result is altered quantitatively in $3+1$ dimensions remains unclear.
\par
In this work, we study quark-hadron duality and its violation for heavy meson mixings in the 't Hooft model.\footnote{Duality violation in the $D^0-\bar{D^0}$ mixing was concerned in Ref.~\cite{Bigi:2000wn}, where the matrix element of the higher dimensional operator that linearly depends on strange quark mass avoiding the strong GIM cancellation was mainly discussed.}
We first calculate the meson mixings based on the box diagrams in two-dimensions, corresponding to the contributions of four-quark operators in the HQE. Also calculated is the same observable based on the exclusive sum over final states, where the two-body decays are dominant in the large-$N_c$ limit since $n$-mesons' coupling is suppressed by $N_c^{1-n/2}$. To perform the exclusive analysis, by following the formalism in Refs.~\cite{Grinstein:1997xk, Grinstein:1998gc}, we represent the topological amplitude \cite{Chau:1982da, Chau:1986jb, Chau:1987tk, Chau:1989tk} in terms of the overlap integrals for meson wave functions, which can be determined as numerical solutions to the 't Hooft equation. Then, the two calculated quantities are compared, in order to check the validity of the HQE. A non-trivial point in this comparison is that the GIM mechanism potentially affects the order of magnitudes of the observables. The investigation for the $D^0-\bar{D^0}$ mixing, subject to the strong GIM cancellation, is distinguished from ones for the $B^0_q-\bar{B}^0_q (q=d, s)$ mixing. We show that the a large correction to the box diagram is realized for $D^0-\bar{D^0}$ when the phase space function is given solely by 4D-like one with certain choices of strange quark mass. As for the $B_q^0-\bar{B}_q^0$ mixing, the correction is much smaller than that for the $D^0-\bar{D}^0$ mixing, and consistent with the realistic observations in four-dimensions. Furthermore, this work deals with heavy meson decays into light mesons, such as $D^0\to \pi^+\pi^-\to \bar{D^0}$, in addition to decays into heavy mesons.  Little has been known for duality in the former case while for latter, especially $\bar{B}^0_s\to D_s^{(*)}D_s^{(*)}\to B^0_s$, an agreement between the partonic rate and the exclusive rate is shown \cite{Aleksan:1993qp} (see also the later study \cite{Chua:2011er}) in the small-velocity limit \cite{Shifman:1987rj} together with heavy quark and large-$N_c$ limits. 
\par
This paper is organized as follows: In Sec.~\ref{Sec:2}, the formalism of the meson mixings, including formulae of the width differences, is exhibited. In Sec.~\ref{Sec:3}, we first recapitulate the 't Hooft model to establish the notation. Subsequently calculated is the absorptive part of partonic transitions, $c\bar{u}\to u\bar{c}$ and $b\bar{q}\to q\bar{b}$ ($q=d, s$). Then, by taking the matrix elements, we obtain the formula of the HQE from the four-quark operators. The counterpart in the exclusive approach is also obtained in the large-$N_c$ limit. We show the numerical results in regards to violation of local duality in Sec.~\ref{Sec:4}, by first analyzing the width differences from the individual flavors and then showing the results in the presence of the GIM mechanism. Finally, we conclude in Sec.~\ref{Sec:5}.
\section{Formalism in the CP conserving limit}
\label{Sec:2}
\subsection{$D^0-\bar{D^0}$, $B^0_d-\bar{B^0_d}$  and $B^0_s-\bar{B^0_s}$ mixings}\label{Sec:2.1}\noindent
For the the $D^0-\bar{D^0}$ mixing, we introduce mass eigenstates denoted by $\ket{D_{1, 2}}$ that diagonalize the Schr\"{o}dinger equations \cite{Zyla:2020zbs} in the CP-conserving limit, where $\ket{D_1} (\ket{D_2})$ coincides with a CP-even (odd) state. The off-diagonal element of the mixing matrix is given by,
\bea
M_{21}^{(D^0)}-\frac{i}{2}\Gamma_{21}^{(D^0)}=\frac{\bra{\bar{D^0}}\mathcal{H}_{W}^{(D^0)}\ket{D^0}}{2M_{D^0}},\quad
\mathcal{H}_{W}^{(D^0)}=\mathcal{H}_{W}^{(D^0,\:\mathrm{dis})}-\frac{i}{2}\mathcal{H}_{W}^{(D^0,\:\mathrm{abs})}.
\label{Eq:G12def1}
\eea
$M_{21}^{(D^0)}$ and $\Gamma_{21}^{(D^0)}$ are associated with the contributions of off-shell and on-shell intermediate states, respectively. The width difference between the two CP states defined by $\Delta \Gamma_D=\Gamma_1^{(D^0)}-\Gamma_2^{(D^0)}$ can be expressed in terms of the off-diagonal element of the mixing matrix,
\bea
\Delta \Gamma_D=2\Gamma_{21}^{(D^0)},\label{Eq:DelMG}
\eea
in the CP-conserving limit. The sign of the above observable is to be determined experimentally in this convention. 
\par
As for the $B^0_q-\bar{B^0_q}$ mixing ($q=d, s$), a commonly adopted convention is based on $\ket{B_H}$ and $\ket{B_L}$, heavier and lighter eigenstates. In the CP conserving limit, one finds that the sign of $\Delta \Gamma=\Gamma_H-\Gamma_L$ depends on that of $M_{12}$ unlike in Eq.~(\ref{Eq:DelMG}), as can be seen in Eq.~(2.16) of Ref.~\cite{Buras:1984pq}. In order to compare the results of the $D^0-\bar{D^0}$ and the $B^0_q-\bar{B^0_q}$ mixings on the equal footing, the convention similar to that of the $D^0-\bar{D^0}$ mixing is adopted in the $B^0_{q}-\bar{B^0_{q}}$ mixing. That is, we introduce mass eigenstates of $\ket{B_{1, 2}}$, where $\ket{B_1} (\ket{B_2})$ is a CP-even (odd) state, and define $\Delta \Gamma_{B_q}=\Gamma_1^{(B_q)}-\Gamma_2^{(B_q)}$. For this case, the following notation similar to one for the $D^0-\bar{D^0}$ mixing is introduced,
\bea
&M_{12}^{(\bar{B^0_q})}-\displaystyle\frac{i}{2}\Gamma_{12}^{(\bar{B^0_q})}=\displaystyle\frac{\bra{B^0_q}\mathcal{H}_{W}^{\bar{(B^0_q)}}\ket{\bar{B^0_q}}}{2M_{B^0_q}},\quad\mathcal{H}_{W}^{(\bar{B^0_q})}=\mathcal{H}_{W}^{(\bar{B^0_q},\:\mathrm{dis})}-\frac{i}{2}\mathcal{H}_{W}^{(\bar{B^0_q},\:\mathrm{abs})},&\label{Eq:BBbar}\\
&\Delta \Gamma_{B_{q}}=2\Gamma_{12}^{(\bar{B^0_q})}.&\label{Eq:DelMGforB}
\eea
Hereafter we exploit $\Gamma_{21}^{(D^0)}=\Gamma_{12}^{(D^0)}$, valid in the CP conserving limit, and do not utilize the notation of $\Gamma_{21}^{(D^0)}$ for brevity: we calculate the $D^0\to \bar{D^0}$ transition for the $D^0-\bar{D^0}$ mixing while $\bar{B^0_q}\to B^0_q$ is computed for the $B^0_{q}-\bar{B^0_{q}}$ mixing, in the common notation of $\Gamma_{12}$.
\subsection{Width differences}
\label{Sec:2.2}\noindent
For the $D^0-\bar{D^0}$ and $B^0_q-\bar{B^0_q}$ mixings, $\Gamma_{12}$ in Eqs.~(\ref{Eq:G12def1}, \ref{Eq:BBbar}) are given by the following expressions $(\alpha = \mathrm{inc}, \mathrm{exc})$,
\bea
\Gamma_{12}^{(D^0,\:\mathrm{\alpha})}&=&\lambda_d^2\Gamma^{(D^0,\:\mathrm{\alpha})}_{dd}+2\lambda_s \lambda_d\Gamma^{(D^0,\:\mathrm{\alpha})}_{sd}+\lambda_s^2\Gamma^{(D^0,\:\mathrm{\alpha})}_{ss}\label{Eq:incdefD}
,\\
\Gamma_{12}^{(\bar{B^0_q},\:\mathrm{\alpha})}&=&\lambda_{u(q)}^2\Gamma^{(\bar{B^0_q},\:\mathrm{\alpha})}_{uu}+2\lambda_{c(q)} \lambda_{u(q)}\Gamma^{(\bar{B^0_q},\:\mathrm{\alpha})}_{cu}+\lambda_{c(q)}^2\Gamma^{(\bar{B^0_q},\:\mathrm{\alpha})}_{cc}.\label{Eq:incdefB}
\quad 
\eea
The products of the CKM matrix elements are defined by,
\bea
\lambda_i&=&V_{ci}^*V_{ui},\quad (i=d, s, b)\label{Eq:lami}\\
\lambda_{j(q)}&=&V_{jb}V_{jq}^*,\quad (j=u, c, t~\mathrm{and}~q=d, s)\label{Eq:lamq}
\eea
where in the CP conserving limit, $\lambda_i$ and  $\lambda_{j(q)}$ are both real-valued. 
We shall adopt the Wolfenstein parameters of Particle Data Group (PDG) \cite{Zyla:2020zbs} to calculate Eqs.~(\ref{Eq:lami}, \ref{Eq:lamq}) for the numerical results presented in Sec.~\ref{Sec:4.2}. $\Gamma_{12}^{(H,\:\mathrm{inc})} (H=D^0, \bar{B^0_d}, \bar{B^0_s})$ is evaluated through the quark-level analysis of HQE while $\Gamma_{12}^{(H,\:\mathrm{exc})}$ is computed on the basis of the solution to the 't Hooft equation by taking sum over exclusive hadronic final states. The three pieces, $\Gamma_{dd}^{(D^0,\:\mathrm{inc})}, \Gamma^{(D^0,\:\mathrm{inc})}_{sd}$ and $\Gamma^{(D^0,\:\mathrm{inc})}_{ss}$ (and similar objects for $\bar{B^0_q}$), represent individual quark contributions in the loop while the intermediate particles are given by the associated bound states for ones with $\mathrm{inc}\to\mathrm{exc}$.
\par
Exploiting the unitarity relation, $\lambda_d+\lambda_s+\lambda_b=0~(\lambda_{u(q)}+\lambda_{c(q)}+\lambda_{t(q)}=0)$, one can eliminate $\lambda_d~(\lambda_{u(q)})$ in Eq. (\ref{Eq:incdefD}) (Eq. (\ref{Eq:incdefB})) and write,
\bea
\Gamma_{12}^{(D^0,\:\mathrm{\alpha})}&=&\lambda_s^2\Gamma_{(\mathrm{GIM},~1)}^{(D^0,~\alpha)}+2\lambda_s\lambda_b\Gamma_{(\mathrm{GIM},~2)}^{(D^0,~\alpha)}+\lambda_b^2\Gamma^{(D^0,\:\mathrm{\alpha})}_{dd},\label{Eq:G12incD}\quad\qquad\\
\Gamma_{12}^{(\bar{B^0_q},\:\mathrm{\alpha})}&=&\lambda_{c(q)}^2\Gamma_{(\mathrm{GIM},~1)}^{(\bar{B^0_q},~\alpha)}+2\lambda_{c(q)}\lambda_{t(q)}\Gamma_{(\mathrm{GIM},~2)}^{(\bar{B^0_q},~\alpha)}+\lambda_{t(q)}^2\Gamma^{(\bar{B^0_q},\:\mathrm{\alpha})}_{uu},\quad\label{Eq:G12incB}\qquad
\eea
where the combinations for individual contributions of flavors are given by,
\bea
\Gamma_{(\mathrm{GIM},~1)}^{(D^0,~\alpha)}&=&\Gamma^{(D^0,\:\mathrm{\alpha})}_{dd}+\Gamma^{(D^0,\:\mathrm{\alpha})}_{ss}-2\Gamma^{(D^0,\:\mathrm{\alpha})}_{sd}\label{Eq:GIM1}\\
\Gamma_{(\mathrm{GIM},~2)}^{(D^0,~\alpha)}&=&\Gamma^{(D^0,\:\mathrm{\alpha})}_{dd}-\Gamma^{(D^0,\:\mathrm{\alpha})}_{sd}\label{Eq:GIM2}\\
\Gamma_{(\mathrm{GIM},~1)}^{(\bar{B^0_q},~\alpha)}&=&\Gamma^{(\bar{B^0_q},\:\mathrm{\alpha})}_{uu}+\Gamma^{(\bar{B^0_q},\:\mathrm{\alpha})}_{cc}-2\Gamma^{(\bar{B^0_q},\:\mathrm{\alpha})}_{cu}\label{Eq:GIM3}\\
\Gamma_{(\mathrm{GIM},~2)}^{(\bar{B^0_q},~\alpha)}&=&\Gamma^{(\bar{B^0_q},\:\mathrm{\alpha})}_{uu}-\Gamma^{(\bar{B^0_q},\:\mathrm{\alpha})}_{cu}.\label{Eq:GIM4}
\eea
One finds that Eqs.~(\ref{Eq:GIM1}, \ref{Eq:GIM2}) and Eqs.~(\ref{Eq:GIM3}, \ref{Eq:GIM4}) vanish for $s=d$ and $c=u$, respectively, so that the first two terms in Eq.~(\ref{Eq:G12incD}) and Eq.~(\ref{Eq:G12incB}) are sensitive to flavor symmetry breakings.
\par
The characteristic differences between $D^0-\bar{D^0}$, $B_d^0-\bar{B_d^0}$ and $B_s^0-\bar{B_s^0}$ mixings can be found in Eqs.~(\ref{Eq:incdefD}-\ref{Eq:GIM4}). To see this, we exploit the hierarchy of the CKM matrix elements, $|\lambda_s|\gg  |\lambda_b|$ for the $D^0-\bar{D^0}$ mixing and $|\lambda_{u(s)}|\ll  |\lambda_{c(s)}|$ for the $B_s^0-\bar{B_s^0}$ mixing. If the SU(3) breaking in Eq.~(\ref{Eq:GIM1}) is larger than the suppression from $\lambda_b$ for $D^0-\bar{D^0}$ mixing, we find that the $D^0-\bar{D^0}$ and $B^0_s-\bar{B^0_s}$ mixings are approximated by one term,
\bea
\Gamma_{12}^{(D^0)}&\simeq&\lambda_s^2\Gamma_{(\mathrm{GIM},~1)}^{(D^0)},\label{Eq:Appro1}\\
\Gamma_{12}^{(\bar{B^0_s})}&\simeq&\lambda_{c(s)}^2\Gamma_{cc}^{(B^0_s)},\label{Eq:Appro2}
\eea
where Eq.~(\ref{Eq:G12incD}) is used for Eq.~(\ref{Eq:Appro1}) while Eq.~(\ref{Eq:incdefB}) is considered for Eq.~(\ref{Eq:Appro2}). As for the $B_d^0-\bar{B_d^0}$ mixing, $|\lambda_{u(d)}|,  |\lambda_{c(d)}|$ and $|\lambda_{t(d)}|$ are comparable so that the formula corresponding to Eqs.~(\ref{Eq:Appro1}, \ref{Eq:Appro2}) is not simplified, yet the strong sensitivity to (GIM, 1) is absent. It should be stressed that the order of magnitude for $\Gamma_{12}$ is characterized by flavor symmetry breaking specifically in the case of the $D^0-\bar{D^0}$ mixing, to be contrasted with the case of the $B_q^0-\bar{B}_q^0$ mixing. This aspect, arising from the different CKM structures in $D, B_d, B_s$ systems, affects the order of the magnitude of final results, as we shall see in Sec.~\ref{Sec:4.2}.
In the later numerical analysis for violation of local duality, we use the exact formulas in Eqs.~(\ref{Eq:G12incD}-\ref{Eq:GIM4}) instead of Eqs.~(\ref{Eq:Appro1}, \ref{Eq:Appro2}).
\par
In the CP conserving limit, $\Gamma_{12}^{(H,\: \mathrm{exc})}$ is expressed as a sum over final states \cite{Falk:2001hx, Cheng:2010rv} for $(H, \bar{H})=(D^0, \bar{D^0}), (\bar{B^0_d}, B^0_d)$ and $(\bar{B^0_s}, B^0_s)$,
\bea
\Gamma_{12}^{(H,\: \mathrm{exc})}
&=&\frac{1}{2}\displaystyle\sum_n\rho_n\left(\bra{\bar{H}}\mathcal{H}_W^{|\Delta F|=1}\ket{n}\bra{n}\mathcal{H}_W^{|\Delta F|=1}\ket{H}\right.\nn\\\ &&\left.+\bra{H}\mathcal{H}_W^{|\Delta F|=1}\ket{n}\bra{n}\mathcal{H}_W^{|\Delta F|=1}\ket{\bar{H}}\right).\label{Eq:formpre}
\eea
with $\rho_n$ being the phase space factor. By using CP transform, one rewrites the above formula,
\bea
\Gamma_{12}^{(H,\: \mathrm{exc})}&=&\frac{1}{2}\displaystyle\sum_n\eta(n)\rho_n\left(\bra{H}\mathcal{H}_W^{|\Delta F|=1}\ket{\bar{n}}\bra{n}\mathcal{H}_W^{|\Delta F|=1}\ket{H}\right.\nn\\\ &&\left.+\bra{H}\mathcal{H}_W^{|\Delta F|=1}\ket{n}\bra{\bar{n}}\mathcal{H}_W^{|\Delta F|=1}\ket{H}\right).\label{Eq:G12def2}
\eea
where $\eta(n)$ is a phase that depends on each intermediate state.
\section{Inclusive and exclusive analyses in $1+1$ dimensions}
\label{Sec:3}
\subsection{The 't Hooft model}
\label{Sec:3.1}\noindent
The QCD Lagrangian in $1+1$ dimensions has a form apparently similar to one in $3+1$ dimensions,
\bea
\mathcal{L}&=&-\frac{1}{4}G_{\mu\nu}^a G^{\mu\nu}_a+\displaystyle\sum_f\bar{\psi}_f (i\slashed{D}-m_f)\psi_f,
\eea
with the covariant derivative defined by $iD_\mu =i\partial_\mu + g A_\mu$. For the second term, the sum runs over flavors. $m_f$ and $g$ are a bare mass and a bare coupling, respectively, both of which have a unit mass dimension in $1+1$ spacetime. We introduce the following notation of the QCD coupling,
\bea
\beta^2 = \frac{g^2N_c}{2\pi}.
\eea
$\beta$ is require to be a constant in the large-$N_c$ limit for the sensible counting for $N_c$, and gives a unit for any dimensional quantities in the model. We adopt the light-cone gauge satisfying $A_-=0$, in which case the theory becomes ghost-free while the field strength is simplified to be effectively Abelian. With the notations introduced above, the 't Hooft equation is given by,
\bea
M_n^2\phi_n^{q_1\bar{q}_2}(x)=\left(\frac{m_1^2-\beta^2}{x}+\frac{m_2^2-\beta^2}{1-x}\right)\phi_n^{(q_1\bar{q}_2)}-\beta^2\: \mathrm{Pr}\int_0^1\D y\frac{\phi_n^{(q_1\bar{q}_2)}(y)}{(x-y)^2}.\label{Eq:tHooft}
\eea
where $x$ and $1-x$ represent the light-cone momentum fractions that are carried by $q_1$ and $\bar{q}_2$, respectively. $M_n$ denotes the meson mass while $m_1$ and $m_2$ are bare masses of $q_1$ and $\bar{q}_2$, respectively. $\phi_n$ is a meson wave function of the $n$-th $(n=0, 1, \cdots)$ radial state that satisfies the boundary conditions, $\phi_n(0)=\phi_n(1)=0$. States labeled by $n=$ even are pseudoscalar mesons with $n=0$ being the ground state, the lightest hadron. The other states with $n=$ odd are scalar mesons. As was shown by 't Hooft, Eq.~(\ref{Eq:tHooft}) is independent of the infrared cut-off. The renormalizations for fermion masses were already taken into account by shifting the bare masses, $m_1^2\to m_1^2-\beta^2$ and $m_2^2\to m_2^2-\beta^2$, in Eq.~(\ref{Eq:tHooft}). Furthermore, by introducing a meson decay constant for the $n$-th  radial state consisting of $q_1$ and $\bar{q_2}$,
\bea
f_n^{(q_1\bar{q_2})}&=&\sqrt{\frac{N_c}{\pi}}c_n^{(q_1\bar{q_2})},\label{Eq:decayc1}\\
c_n^{(q_1\bar{q_2})}&=&\int_{0}^{1}\D x \phi_n^{(q_1\bar{q}_2)}(x),\label{Eq:decayc2}
\eea
one writes a matrix element for the axial current,
\bea
\bra{0}\bar{q_2}\gamma_\mu \gamma_5 q_1 \ket{H(p)}&=&f_Hp_\mu. \label{Eq:Ve}
\eea
Above we used the mesonic notation, $H$, for the ground state consisting of $q_1\bar{q_2}$, corresponding to $n=0$ in Eq.~(\ref{Eq:decayc1}, \ref{Eq:decayc2}). The matrix element of the pseudoscalar bilinear similar to Eq.~(\ref{Eq:Ve}) can be derived by using the equation of motion while one for the scalar bilinear vanishes. As for the matrix element of the vector current, it can be rewritten as one for the axial vector current in Eq.~(\ref{Eq:Ve}) by using the relation of the gamma matrix in two-dimensions, as is done in Appendix.
\subsection{HQE from leading operators}
\label{Sec:3.2}\noindent
\begin{figure}[t]
\begin{center}
\subfigure[]{\includegraphics[width=0.2\columnwidth]{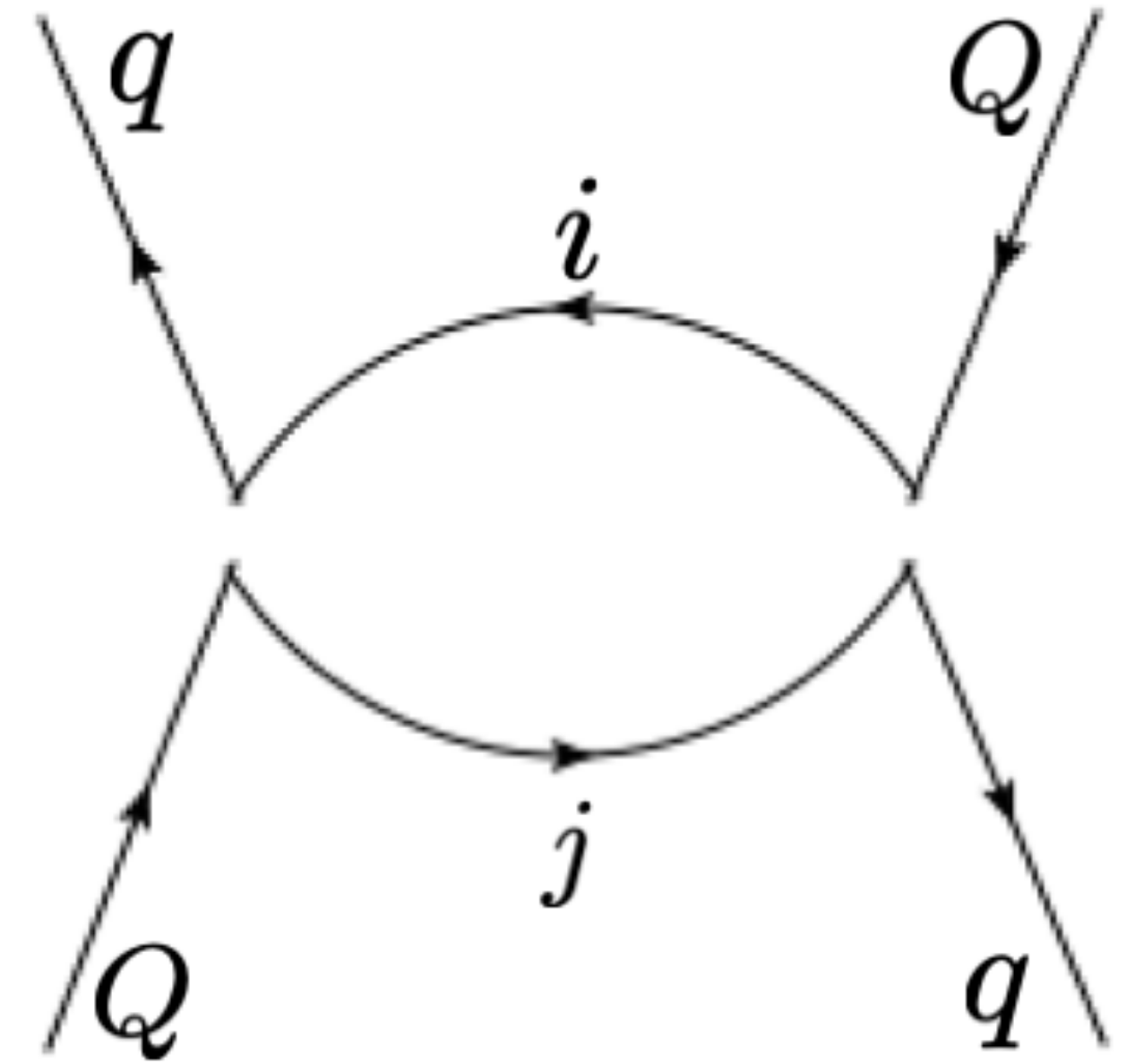}}\hspace{10mm}
\subfigure[]{\includegraphics[width=0.2\columnwidth]{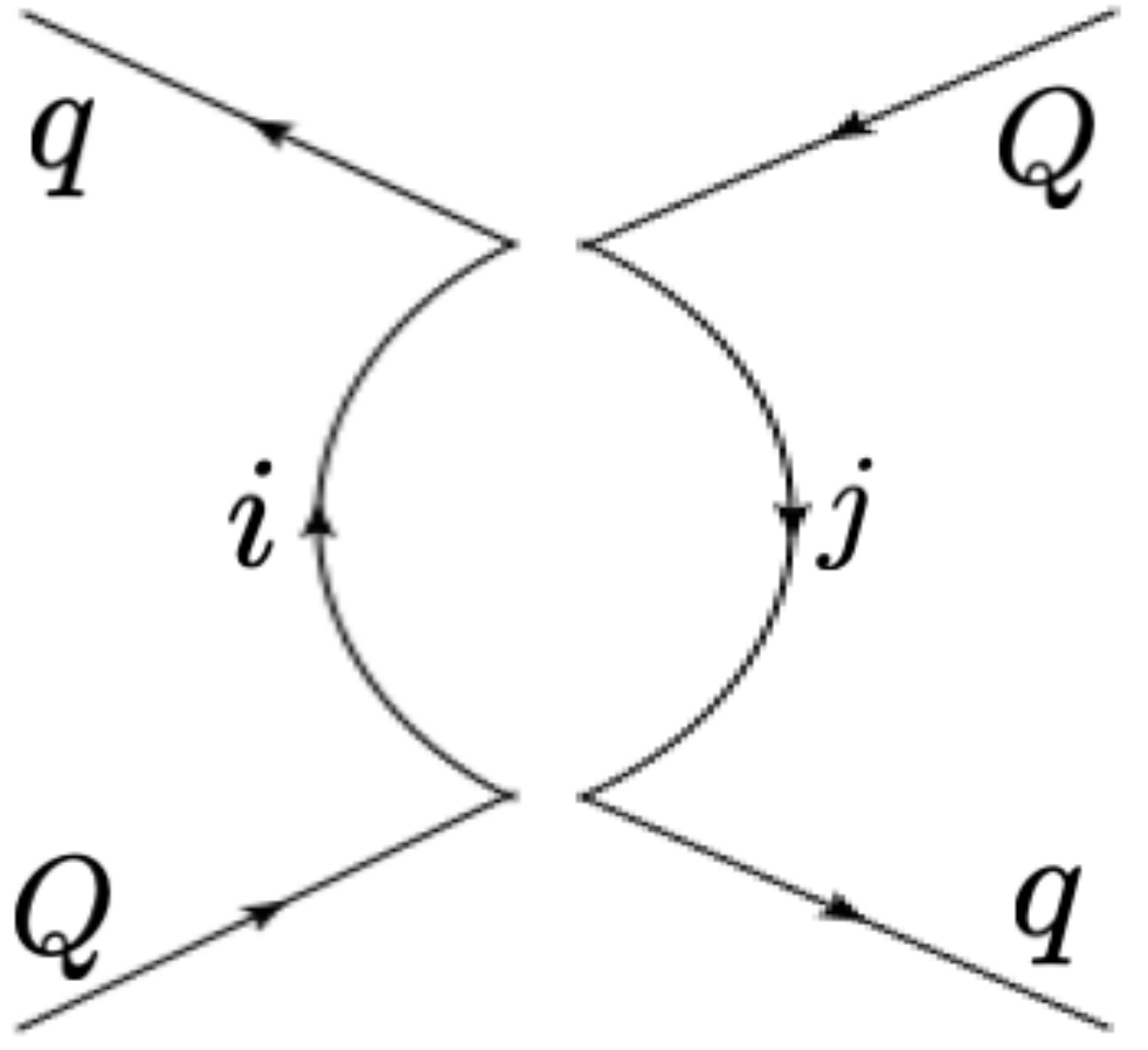}}
\end{center}
\vspace{-5mm}
\caption{Partonic processes of $Q\bar{q}\to q\bar{Q}$ with $(Q, \bar{q})$ being $(c, \bar{u})$ or $(b, \bar{d})$. $i$ and $j$ denote down-type (up-type) quarks for the former (latter). Only shown are color flows, and the $W$ bosons are omitted.}
\label{Fig:1}
\end{figure}
We consider the weak vertex that has a generalized Lorentz structure parameterized as,
\bea
\frac{-ig_2}{\sqrt{2}}V_{\mathrm{CKM}}\gamma^\mu (c_\mathrm{V}+c_\mathrm{A}\gamma_5),\label{Eq:Lorentz}
\eea
with $V_{\mathrm{CKM}}$ being the CKM matrix element associated with a given process. $c_\mathrm{V}=-c_\mathrm{A}=1/2$ corresponds to the case where the weak interaction proceeds via the standard model-like V$-$A current. The $W$ boson propagator given as in $3+1$ dimensions is,
\bea
\frac{-i}{q^2-M_W^2+i\epsilon}\left(g_{\mu\nu}-\xi \frac{q_\mu q_\nu}{M_W^2}\right),\label{Eq:prop}
\eea
where fixing $\xi=1$ leads to the unitary gauge, in which case the contributions of the charged-Goldstone bosons are absent. We keep the contribution that is dominant in the limit of $M_{W}\to\infty$, corresponding to the $g_{\mu\nu}$ part in Eq.~(\ref{Eq:prop}). Below, by using these Feynman rules, we give the effective Hamiltonian leading to the absorptive parts of $Q\bar{q}\to q\bar{Q}$ transition with $(Q, \bar{q})$ being $(c, \bar{u})$, $(b, \bar{d})$ or $(b, \bar{s})$ shown in Fig.~\ref{Fig:1}. The detail of the calculation is given in Appendix. As a result, the absorptive parts of the effective Hamiltonian that contribute to the heavy meson mixing in the considered approximations are given by,
\bea
\mathcal{H}_W^{(H,\:\mathrm{abs})}&=&\displaystyle\sum_{i, j}\lambda_i\lambda_j(C^{\rm A}_{ij}\mathcal{O}^{\rm A}+C^{\rm P}_{ij}\mathcal{O}^{\rm P}),\label{Eq:Hham}
\eea
The coefficients and the four-quark operators are given by,
\bea
C^{\rm A}_{ij}&=&+4G_F^2(c_{\rm V}^2-c_{\rm A}^2)
\left[(c_{\rm V}^2-c_{\rm A}^2)\left(F^{\rm (th)}_{ij} + 2G^{\rm (th)}_{ij}\right)-(c_{\rm V}^2+c_{\rm A}^2)\left(I^{\rm (th)}_{ij}+I^{\rm (th)}_{ji}\right)\right],\label{Eq:CA}\qquad\\
C^{\rm P}_{ij}&=&-4G_F^2(c_{\rm V}^2-c_{\rm A}^2)
\left[(c_{\rm V}^2-c_{\rm A}^2)\left(G^{\rm (th)}_{ij}+2H^{\rm (th)}_{ij}\right)+(c_{\rm V}^2+c_{\rm A}^2)\left(I^{\rm (th)}_{ij}+I^{\rm (th)}_{ji}\right)\right],\label{Eq:CP}\qquad\\
\mathcal{O}^{\rm A} &=&(\bar{q}^\alpha\gamma^\mu \gamma_5Q^\alpha)(\bar{q}^\beta\gamma_\mu \gamma_5 Q^\beta),\\
\mathcal{O}^{\rm P} &=&(\bar{q}^\alpha i\gamma_5Q^\alpha)(\bar{q}^\beta i \gamma_5 Q^\beta).
\eea
Here $F^{\rm (th)}_{ij}, G^{\rm (th)}_{ij}, H^{\rm (th)}_{ij}$ and $I^{\rm (th)}_{ij}$ represent phase space functions that have non-zero values in a physical region,
\bea
F^{\rm (th)}_{ij}&=&\sqrt{1-2(z_i+z_j)+(z_i-z_j)^2},\label{Eq:phasespace0}\\
G^{\rm (th)}_{ij}&=&\frac{z_i+z_j-(z_i-z_j)^2}{\sqrt{1-2(z_i+z_j)+(z_i-z_j)^2}},\label{Eq:phasespaceG}\\
H^{\rm (th)}_{ij}&=&\frac{\sqrt{z_iz_j}}{\sqrt{1-2(z_i+z_j)+(z_i-z_j)^2}},\label{Eq:phasespaceH}\\
I^{\rm (th)}_{ij}&=&\frac{\sqrt{z_i}(1+z_i-z_j)}{\sqrt{1-2(z_i+z_j)+(z_i-z_j)^2}},
\label{Eq:phasespace}
\eea
with $z_\beta=m_\beta^2/m_Q^2~(\beta=i, j)$.  The leading contribution for large $m_Q$ solely comes from the term proportional to $F^{\rm (th)}_{ij}$.
One finds that the coefficients given in Eqs.~(\ref{Eq:CA}, \ref{Eq:CP}) are proportional to $(c_{\rm V}^2 - c_{\rm A}^2)$. Hence, the observables in meson mixings for $c_{\rm V}=\pm c_{\rm A}$ corresponding to the $V\pm A$ current vanish, which is not seen in four-dimensions. This is partially attributed to the fact that either vector current or axial current is reducible and can be written by another in two-dimensions. The derivation for Eqs.~(\ref{Eq:CA}, \ref{Eq:CP}) by means of the Fiertz rearrangements in two-dimensions is given in Appendix. Furthermore, the non-vanishing result in the limit of $m_i, m_j\to 0$ for the $V\times V$ current ($c_\mathrm{V}\neq 0, c_\mathrm{A}=0$) observed via Eqs.~(\ref{Eq:CA}, \ref{Eq:CP}) is to be contrasted with Ref.~\cite{Bigi:1998kc}, where the contribution of the four-fermion operator in the annihilation-topology, calculated as an absorptive part, is shown to vanish at zeroth order in strong interaction.\par
The matrix elements in Eq.~(\ref{Eq:Hham}) can be taken on the basis of the factorization in the large-$N_c$ limit with Eq.~(\ref{Eq:Ve}),
\bea
\frac{\bra{\bar{H}}\mathcal{O}_{\rm A}\ket{H}}{2M_{H}}&=&f_H^2 M_H,\label{Eq:fact0}\\
\frac{\bra{\bar{H}}\mathcal{O}_{\rm P}\ket{H}}{2M_H}&=&f_H^2 M_HR.
\label{Eq:fact}
\eea
with $R=[M_H/(m_Q+m_q)]^2$. On r.h.s. of Eqs.~(\ref{Eq:fact0}, \ref{Eq:fact}), the factor two, arising from two possible ways for taking the currents in inserting vacuum, are considered, and is cancelled out with $1/2$ on l.h.s. If we go beyond large-$N_c$ limit, an evaluation the non-perturbative matrix elements in Eqs.~(\ref{Eq:fact0}, \ref{Eq:fact}) should be made, that is beyond our current scope. As long as the four-quark operators are concerned, however, the matrix elements do not give sources of flavor symmetry breaking in Eqs.~(\ref{Eq:GIM1}-\ref{Eq:GIM4}).
\par
As a main result in this subsection, one finally obtains the HQE expression of the four-quark operators,
\bea
\Gamma^{(H,\:\mathrm{inc})}_{ij}&=&(C_{\rm A} + C_{\rm P}R)f_H^2 M_H\label{Eq:incres}
\eea
where again $H$ is either $D^0, \bar{B^0_d}$ or $\bar{B^0_s}$ and $(i, j)$ runs $(d, d), (s, d), (s, s)$ for the first case and $(u, u), (c, u), (c, c)$ for the latter two cases. In the limit of $m_Q \to \infty$, it is well-know that $c_0^{(Q\bar{q})}\to 1/\sqrt{m_Q}$ and $M_{H}\sim m_Q+\mathcal{O}(m_Q^0)$ follow, so that $\Gamma_{ij}^{(H,\:\mathrm{inc})}$ behaves like $\Gamma_{ij}^{(H,\:\mathrm{inc})}\propto \mathrm{const}.$, to be contrasted with the case in $3+1$ dimensions, $\Gamma_{ij}^{(H,\:\mathrm{inc})}\propto m_Q^2$, as can be seen from Refs.~\cite{Hagelin:1981zk, Cheng:1982hq, Buras:1984pq}. This difference results from the fact that both Fermi constant and decay constant are dimensionless in $1+1$ spacetime. If we take the massless limit of internal quarks, Eq.~(\ref{Eq:incres}) is recast into,
\bea
\Gamma^{(H,\:\mathrm{inc})}_{ij}&\to&4G_F^2(c_\mathrm{V}^2-c_\mathrm{A}^2)^2f_{H}^2M_{H}.\label{Eq:incres2}
\eea
As we shall see later, Eq.~(\ref{Eq:incres2}) agrees with the exclusive result in the same limit.
\par
The $1/m_Q$ expansion of the contributions of the four-quark operators in Eq.~(\ref{Eq:incres}) can be readily studied in the static limit, $m_Q=m_1\to \infty$, in Eq.~(\ref{Eq:tHooft}) as was first discussed in Refs.~\cite{Burkardt:1991ea,Burkardt:1992qm} with $t=(1-x)m_Q$ and $\psi_n(t)=\phi_n(1-t/m_Q)/\sqrt{m_Q}$. Below, we give the final results for the ground state in Ref.~\cite{Lebed:2000gm},
\bea
c_0^{(Q\bar{q})}\sqrt{m_Q}&=&\left[1-\frac{2}{3}\frac{2\bar{\Lambda}-m_q}{m_Q}\right]F^{(0)}+\mathcal{O}\left(\frac{1}{m_Q^2}\right),\label{Eq:decasym}\\
M_H-m_Q&=&\bar{\Lambda}+\frac{\braket{\bar{Q}(i\vec{D})^2Q}-\beta^2}{2m_Q}+\mathcal{O}\left(\frac{1}{m_Q^2}\right),\label{Eq:lameq}
\eea
where $F^{(n)}$ is a finite object in the static limit,
\bea
F^{(n)}=\int_{0}^{\infty}\mathrm{d}t~\psi_n(t)=\lim_{m_Q\to\infty}c_n^{(Q\bar{q})}\sqrt{m_Q}.
\eea
Moreover, it might be useful to introduce $\delta\equiv R-1$, a quantity power-suppressed by $m_Q$,
where the expansion of $\delta$ is obtained from Eq.~(\ref{Eq:lameq}),
\bea
\delta =2\frac{\bar{\Lambda}-m_q}{m_Q}+\mathcal{O}\left(\frac{1}{m_Q^2}\right).
\eea
One also finds that the phase space functions in Eqs.~(\ref{Eq:phasespace0}-\ref{Eq:phasespace}) give corrections of the $1/m_Q$ expansion due to the expansion formulae,
\bea
F^{\rm (th)}_{ij}&=&1-(z_i+z_j)-2z_iz_j +\mathcal{O}(z^3),\\
G^{\rm (th)}_{ij}&=&(z_i+z_j)+4z_iz_j+\mathcal{O}(z^3),\\
H^{\rm (th)}_{ij}&=&\sqrt{z_iz_j}[1+(z_i+z_j)]+\mathcal{O}(z^3),\\
I^{\rm (th)}_{ij}&=&\sqrt{z_i}[1+2(1+2z_j)z_i + 2z_i^2 +\mathcal{O}(z^3)].
\label{Eq:phase2}
\eea
Only $F^{\rm (th)}_{ij}$ is non-vanishing in the static limit ($z_i, z_j\to 0$) while $G^{\rm (th)}_{ij}, H^{\rm (th)}_{ij}$ and $I^{\rm (th)}_{ij}$ are sub-leading functions. Combining Eqs.~(\ref{Eq:decasym}-\ref{Eq:phase2}), one finds that the expansion for the width difference in Eq.~(\ref{Eq:incres}) starts from $1/m_Q$,
\bea
\Gamma^{(H,\:\mathrm{inc})}_{ij}&=&4G_F^2(c_\mathrm{V}^2-c_\mathrm{A}^2)\frac{N_c}{\pi}\left[F^{(0)}\right]^2\left[(c_\mathrm{V}^2-c_\mathrm{A}^2)\left(1-\frac{5\bar{\Lambda}-4m_q}{3m_Q}\right)
\right.\nn\\
&&\left.-2(c_\mathrm{V}^2+c_\mathrm{A}^2)\frac{m_i+m_j}{m_Q}
+\mathcal{O}\left(\frac{1}{m_Q^2}\right)\right].\label{Eq:expanmQ}
\eea
It is possible to numerically obtain the explicit coefficients of each $1/m_Q$ term with a given mass of the spectator quark as in Ref.~\cite{Lebed:2000gm}. However, since the expansion of $1/m_Q$ is not necessary in our current purpose, the numerical results presented in Sec.~\ref{Sec:4}
are based on Eq.~(\ref{Eq:incres2}) instead of Eq.~(\ref{Eq:expanmQ}).
\par
By using Eq.~(\ref{Eq:incres}), one can write analytical expressions for the GIM combinations in the massless limit of $d$ quark, {\it i.e.,} $z_d=0$. First we give the formula of Eqs.~(\ref{Eq:GIM1}, \ref{Eq:GIM2}) for the case where only $F_{ij}^{(\rm th)}$, corresponding to four-dimension-like phase space function, is considered with the other phase space functions, $G_{ij}^{(\rm th)}, H_{ij}^{(\rm th)}$ and $I_{ij}^{(\rm th)}$ being neglected,
\bea
\left.\Gamma^{(D^0,\:\mathrm{inc})}_{(\mathrm{GIM}, 1)}\right|_{4D-\mathrm{like}}&=&\Gamma^{(D^0,\:\mathrm{inc})}_{dd}\times [-2z_s^2+\mathcal{O}(z_s^3)],\label{Eq:expD1}\qquad\qquad\\
\left.\Gamma^{(D^0,\:\mathrm{inc})}_{(\mathrm{GIM}, 2)}\right|_{4D-\mathrm{like}}&=&\Gamma^{(D^0,\:\mathrm{inc})}_{dd}\times z_s,\label{Eq:expD2}
\eea
where $\Gamma^{(D^0,\:\mathrm{inc})}_{dd}$ defined here is given by r.h.s. of Eq.~(\ref{Eq:incres2}). Although $F_{ij}^{(\rm th)}$ is a leading term in the limit of $m_Q\to \infty$, a certain care must be taken since the inclusion of the 2D-specific phase space function affects the resultant counting in $z_s$. We also give alternative expressions that take account of $F_{ij}^{(\mathrm{th})}, G_{ij}^{(\mathrm{th})}$ and $H_{ij}^{(\mathrm{th})}$ without 
$I_{ij}^{(\mathrm{th})}$,
\bea
\left.\Gamma^{(D^0,\:\mathrm{inc})}_{(\mathrm{GIM}, 1)}\right|_{4D+2D}&=&\Gamma^{(D^0,\:\mathrm{inc})}_{dd}\times\left[-2z_s(1+\delta)+\mathcal{O}(z_s^2)
\right],\label{Eq:expD3}\qquad\qquad\\
\left.\Gamma^{(D^0,\:\mathrm{inc})}_{(\mathrm{GIM}, 2)}\right|_{4D+2D}&=&\Gamma^{(D^0,\:\mathrm{inc})}_{dd}\times \delta z_s,\label{Eq:expD4}
\eea
to be contrasted with Eqs.~(\ref{Eq:expD1}, \ref{Eq:expD2}). Equations (\ref{Eq:expD1}-\ref{Eq:expD4}) clearly show that the width difference is suppressed by SU(3) breaking. It is also possible to consider the GIM combinations in the presence of all $F_{ij}^{(\mathrm{th})}, G_{ij}^{(\mathrm{th})}, H_{ij}^{(\mathrm{th})}$ and $I_{ij}^{(\mathrm{th})}$. 
\subsection{Topological amplitude in the large-$N_c$ limit}
\label{Sec:3.3}\noindent
In the remaining part of this section, we aim to obtain $\Gamma_{12}^{(\mathrm{exc})}$ as an exclusive sum by using the wave functions and masses of mesons from the 't Hooft equation. Below, the on-shell intermediate contributions of $H\to f_1 f_2\to \bar{H}$ to the meson mixings are considered with $f_1$ and $f_2$ being either pseudoscalar or scalar. The contributions of two-body decays are completed for this case since hadronic states with non-zero angular momentum, {\it e.g.,} vector and axial vector mesons, are absent in two-dimensions. 
For neutral mesons, the two-body decay amplitudes are characterized by color-allowed tree ($T$), color-suppressed tree ($C$), exchange ($E$), penguin ($P$), penguin annihilation ($PA$) and penguin exchange ($PE$) diagrams. For explicit decomposition via the topological amplitudes, see Ref.~\cite{Cheng:2012xb}. In the naive counting, $T\propto N_c^{1/2}, C, E, P, PA \propto N_c^{-1/2}$ and $PE \propto N_c^{-3/2}$ follow. Even if we take account of resonant contributions for some of the topological amplitudes, that strengthen $N_c$ dependence \cite{Grinstein:1998gc}, the contribution of $T$ to width is still dominant compared with the others in the large-$N_c$ limit. The leading decay amplitudes from two-body pseudoscalar modes
in the large-$N_c$ limit are given by,
\bea
&A[D^0\to \pi^+\pi^-]=V_{ud}V_{cd}^*T_{(c\bar{u})(d, d)}^{(0, 0)},\quad
A[D^0\to \pi^+K^-]=V_{ud}V_{cs}^*T_{(c\bar{u})(d, s)}^{(0, 0)},&\nn\\
&A[D^0\to K^+\pi^-]=V_{us}V_{cd}^*T_{(c\bar{u})(s, d)}^{(0, 0)},\quad
A[D^0\to K^+K^-]=V_{us}V_{cs}^*T_{(c\bar{u})(s, s)}^{(0, 0)},&\label{Eq:ps1}
\eea
where the superscript, $(0, 0)$, represents the ground states in the final particles while the subscripts, $(c\bar{u})$ and $(i, j)=(d, d), (d, s), (s, d), (s, s)$, stand for the flavors in an initial and final states, respectively. Similarly, one introduces the decay amplitudes of $\bar{B^0_d}$ and $\bar{B^0_s}$ from the color-allowed tree diagrams as follows,
\bea
&A[\bar{B^0_d}\to \pi^-\pi^+]=V_{ud}^*V_{ub}T_{(b\bar{d})(u, u)}^{(0, 0)},\quad \nn
A[\bar{B^0_d}\to \pi^-D^+]=V_{ud}^*V_{cb}T_{(b\bar{d})(u, c)}^{(0, 0)},&\\
&A[\bar{B^0_d}\to D^-\pi^+]=V_{cd}^*V_{ub}T_{(b\bar{d})(c, u)}^{(0, 0)},\quad \nn
A[\bar{B^0_d}\to D^-D^+]=V_{cd}^*V_{cb}T_{(b\bar{d})(c, c)}^{(0, 0)},&\\
&A[\bar{B^0_s}\to K^-K^+]=V_{us}^*V_{ub}T_{(b\bar{s})(u, u)}^{(0, 0)},\quad
A[\bar{B^0_s}\to K^-D^+_s]=V_{us}^*V_{cb}T_{(b\bar{s})(u, c)}^{(0, 0)},&\nn\\
&A[\bar{B^0_s}\to D^-_sK^+]=V_{cs}^*V_{ub}T_{(b\bar{s})(c, u)}^{(0, 0)},\quad
A[\bar{B^0_s}\to D^-_sD^+_s]=V_{cs}^*V_{cb}T_{(b\bar{s})(c, c)}^{(0, 0)}.&\label{Eq:ps2}
\eea
We omitted processes that are given by color-allowed tree diagrams but do not contribute to $B^0_q-\bar{B^0_q}$ through the most color-favored topology, {\it e.g.}, $\bar{B_d^0} \to K^-\pi^+ $ and $\bar{B_d^0} \to D^-_s\pi^+ $. For additionally including the contributions of scalar and other excited particles in the final states, one generalizes the notations in~(\ref{Eq:ps1}) and in~(\ref{Eq:ps2}) into,
\bea
A[(c\bar{u})^{(0)}\to (u\bar{i})^{(k)}(j\bar{u})^{(m)}]&=&V_{ui}V_{cj}^*T_{(c\bar{u})(i, j)}^{(k, m)}\label{Eq:Def},\\
A[(b\bar{q})^{(0)}\to (d\bar{i})^{(k)}(j\bar{q})^{(m)}]&=&V_{iq}^*V_{jb}T_{(b\bar{q})(i, j)}^{(k, m)}\label{Eq:Def2}.
\eea
Here, mesonic states are denoted by $(i \bar{j})^{(k)}$ with $i$ and $j$ being flavors forming the bound state and $k$ being a label of radially excited states with $q=d, s$ in Eq.~(\ref{Eq:Def2}). The initial states are assigned with the ground states, $D^0$ and $\bar{B}^0_q$, to describe processes relevant for the meson mixings. If one takes $k=m=0$, the definitions in Eq.~(\ref{Eq:Def}) and  Eq.~(\ref{Eq:Def2}) reduce to ones in (\ref{Eq:ps1}) and (\ref{Eq:ps2}), respectively.
\par
By performing the phase space integral in $1+1$ dimensions, one writes the partial decay width for $H\to f_1f_2$ decays,
\bea
\Gamma&=&\frac{|A[H\to f_1f_2]|^2}{4M_{H}^2|p_{12}|},\label{Eq:pec}\\
|p_{12}|&=&\frac{M_{H}}{2}\sqrt{1-2\frac{M_1^2+M_2^2}{M_{H}^2}+\frac{(M_1^2-M_2^2)^2}{M_{H}^4}}.\label{Eq:mom}
\eea
where $p_{12}$ denotes a momentum of either daughter meson in the rest frame of $H$. The peculiarity of the phase space, showing that the width looks divergent when $M_{H}=M_1+M_2$ or equivalently $p_{12}=0$, is present in Eq.~(\ref{Eq:pec}). This point is obviously distinct from the case with $3+1$ dimensions, $\Gamma\propto |p_{12}|$ due to the phase space, leading to the vanishing width for $p_{12}=0$. In the analytical study \cite{Bigi:1998kc}, it was shown that this singularity is cancelled out with an amplitude in the semi-leptonic decay, in which massless particles are involved.
\par
As mentioned, we consider the rigorous large-$N_c$ limit, where the resonant width associated with strong decays vanishes, in which case the topological amplitude does not develop its imaginary part. One can write the individual internal quark contributions on r.h.s. in Eqs.~(\ref{Eq:G12incD}, \ref{Eq:G12incB}) by allocating $n$ in Eq.~(\ref{Eq:G12def2}) on the basis of the relevant quantum numbers.
\begin{figure}
\centering
\includegraphics[clip,width=0.47\columnwidth]{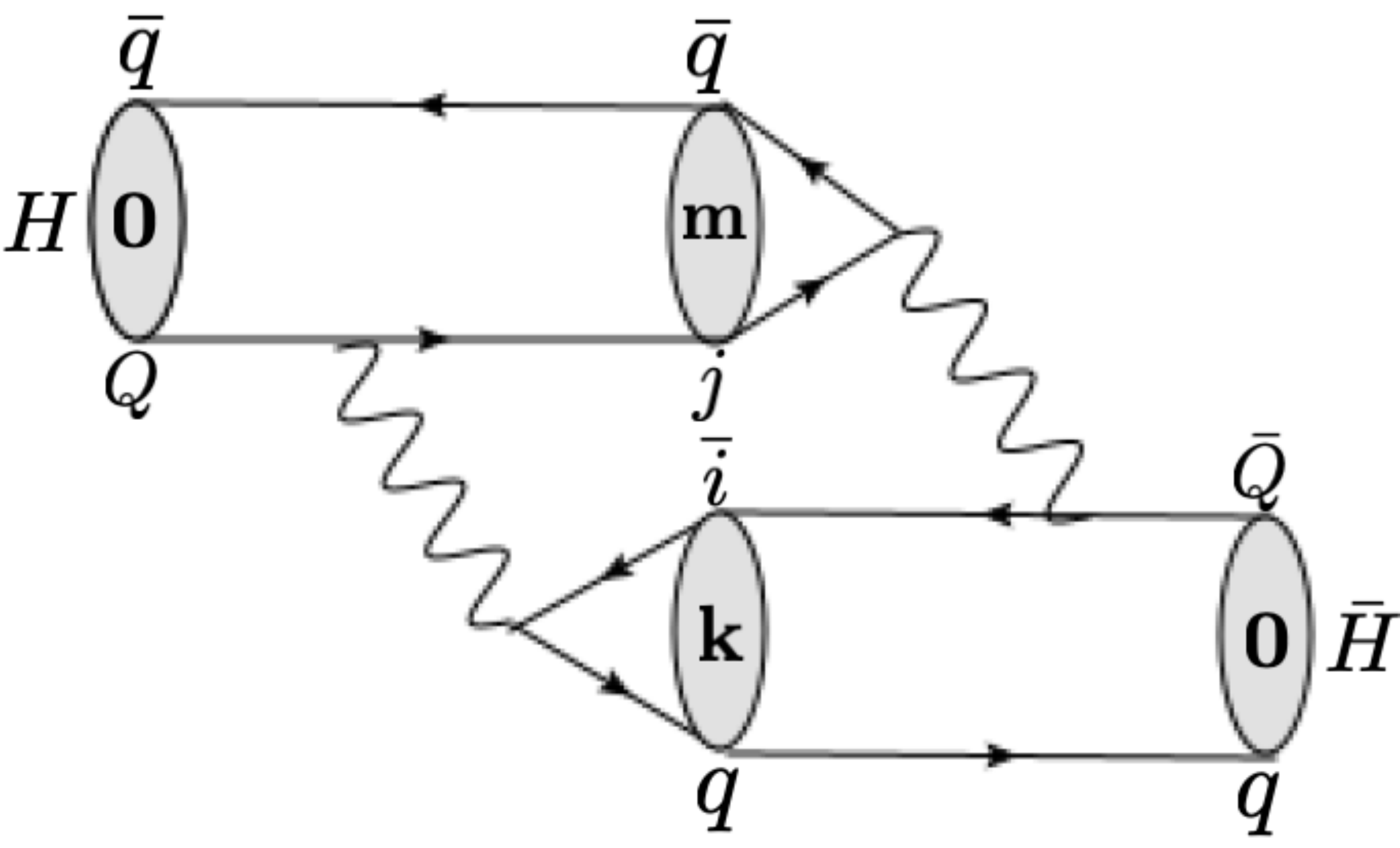}
\caption{Exclusive processes for $H\to \bar{H}$, where $H$ consists of $Q\bar{q}$. $(H, Q, \bar{q})$ is taken as $(D^0, c, \bar{u}), (\bar{B^0_d}, b, \bar{d})$  or $(\bar{B^0_s}, b, \bar{s})$ while the numbers denoted by $\mathbf{0}, \mathbf{k}, \mathbf{m}$ represent labels for the radial states. $i$ and $j$ stand for down-type (up-type) quarks for $H=D^0$ $(\bar{B^0_d}, \bar{B^0_s})$.}
\label{Fig:2}
\end{figure}
The diagrams for exclusive processes, given by the hadronic degrees of freedom in the most color-allowed topology, are shown in Fig.~\ref{Fig:2} for the heavy meson mixings.
Armed with Eq.~(\ref{Eq:G12def2}) and the vanishing of the strong phases, we evaluate Fig.~\ref{Fig:2},
\bea
\Gamma^{(D^0,\:\mathrm{exc})}_{ij}&=&\displaystyle\sum_{k, m}(-1)^{k+m}\frac{T_{(c\bar{u})(i, j)}^{(k, m)} T_{(c\bar{u})(j, i)}^{(m, k)\:*}}{4M_{D^0}^2|p_{km}|}\quad \mathrm{for}~(i, j)=(d, d), (s, d), (s, s),\label{Eq:exc1}\\
\Gamma^{(\bar{B^0_q},\:\mathrm{exc})}_{ij}&=&\displaystyle\sum_{k, m}(-1)^{k+m}\frac{T_{(b\bar{q})(i, j)}^{(k, m)}T_{(b\bar{q})(j, i)}^{(m, k)\:*}}{4M_{B^0_q}^2|p_{km}|}\quad \mathrm{for}~(i, j)=(u, u), (c, u), (c, c).\label{Eq:exc2}
\eea
The momenta denoted by $p_{km}$ is understood as one in Eq.~(\ref{Eq:mom}) with the relevant final state. For the individual $(i, j)$ contribution, the sum over $(k, m)$, representing the tower of kinematically allowed excited particles (in addition to the ground states) for final states, is taken. The prefactor of $(-1)^{k+m}$ comes from $\eta(n)$ in Eq.~(\ref{Eq:G12def2}) to account for the parity-odd property of the topological amplitude in the case of $k+m=\mathrm{even}$ due to the proportionality to the spatial component of a momentum with overall negative signs. It should be also noted that there exists no orbital angular momentum via the relative motion of particles in the final state in two-dimensions.
\par
The explicit formula for color-allowed tree amplitude is obtained in Ref.~\cite{Grinstein:1997xk}. Below we denote the momenta of mesons labeled by $\mathbf{k}, \mathbf{m}$ as $q$ and $p$, respectively. The kinematical variable defined by $\omega=q_-/p_-$ is determined by,
\bea
\omega =\frac{1}{2}\left[1+\left(\frac{q^2-M_m^2}{M_0^2}\right)-\sqrt{1-2\left(\frac{q^2+M_m^2}{M_0^2}\right)+\left(\frac{q^2-M_m^2}{M_0^2}\right)^2}\right].\label{Eq:omega}
\eea
With the generalized Lorentz structure in Eq.~(\ref{Eq:Lorentz}), we can use the formulas \cite{Grinstein:1997xk, Bigi:1999fi}, valid for $(Q, \bar{q})$ equal to $(c, \bar{u}), (b, \bar{d})$ and $(b, \bar{s})$ in the limit of $M_W\to \infty$,
\bea
T^{(k, m)}_{(Q\bar{q})(i, j)}&=&2\sqrt{2}G_F(c_\mathrm{V}^2-c_\mathrm{A}^2)\sqrt{\frac{N_c}{\pi}}c_k^{(q\bar{i})}
\left[\displaystyle\sum_{n=0}\frac{[(-1)^{k}q^2+(-1)^n M_n^2]c_n^{(Q\bar{j})}}{q^2-M_n^2}F_{nm}\right.\nn\\
&&\left.+(-1)^{k+1}q^2\mathcal{C}_m+m_Qm_{j}\mathcal{D}_m\right],
\label{Eq:topo}
\eea
For an on-shell process, $q^2$ is set to $M_k^2$ in Eqs.~(\ref{Eq:omega}, \ref{Eq:topo}). $F_{nm}$ denotes the triple overlap integral while $\mathcal{C}_{m}$ and $\mathcal{D}_{m}$ are the quark-model type contact terms \cite{Bigi:1999fi},
\bea
F_{nm}&=&
\omega(1-\omega)\int_0^1\mathrm{d}x\int_0^1\mathrm{d}y
\frac{\phi_n^{(Q\bar{j})}(x)\phi_m^{(j\bar{q})}(y)}{[\omega(1-x)+(1-\omega)y]^2}\nn\\
&&\times\{\phi_{0}^{(Q\bar{q})}(\omega x)-\phi_{0}^{(Q\bar{q})}[1-(1-\omega)(1-y)]\},\label{Eq:Ftri}\\
\mathcal{C}_m&=&-\frac{1-\omega}{\omega}\int_0^1\mathrm{d}x \phi_0^{(Q\bar{q})}[1-(1-\omega)(1-x)]\phi_m^{(j\bar{q})}(x),\label{Eq:Cdou}\\
\mathcal{D}_m&=&-\omega\int_0^1\mathrm{d}x
\frac{\phi_0^{(Q\bar{q})}[1-(1-\omega)(1-x)]}{1-(1-\omega)(1-x)}
\displaystyle\frac{\phi_m^{(j\bar{q})}(x)}{x},\label{Eq:Ddou}
\eea
\par
One can analytically simplify $m_Q$ dependence of the width difference in Eqs.~(\ref{Eq:exc1},  \ref{Eq:exc2}) in the massless limit of quarks except for heavy quarks in initial and final states. For this case, the only non-vanishing contribution in Fig.~\ref{Fig:2} is $k=m=0$ due to the vanishing property of the decay constants for excited states, $c_n = 0~(n\neq 0)$, for the massless constituents. Since $M_k$ and $M_m$ with $k=m=0$ vanish in this limit, $q^2\to 0$ together with $\omega\to 0$ follows for both of the two interfering amplitudes in Eqs.~(\ref{Eq:exc1},  \ref{Eq:exc2}), in which case the terms except for $\mathcal{C}$ in Eq.~(\ref{Eq:topo}) vanish. Then, the interference of the amplitudes is simplified as,
\bea
T^{(0, 0)}_{(Q\bar{q})(i, j)}T^{(0, 0)\:*}_{(Q\bar{q})(j, i)}&=&8G_F^2(c_\mathrm{V}^2-c_\mathrm{A}^2)^2M_H^4\frac{N_c}{\pi}c_0^{(j\bar{q})}c_0^{(q\bar{i})}\nn\\
&&\times\int_0^1\D x \phi_0^{(Q\bar{q})}(x)\phi_0^{(j\bar{q})}(x)\int_0^1\D y \phi_0^{(Q\bar{q})}(y)\phi_0^{(q\bar{i})}(y).
\eea
By using $c_0=1$ and $\phi_0(x)=1$ (except for the end points) for massless constituents, we find that the width difference in Eqs.~(\ref{Eq:exc1}, \ref{Eq:exc2}) is reduced to,
\bea
\Gamma^{(H,\:\mathrm{exc})}_{ij}&=&4G_F^2(c_{\rm V}^2-c_{\rm A}^2)^2 f_{H}^{2}M_H,\label{Eq:localduality}
\eea
It should be noted that Eq.~(\ref{Eq:localduality}) agrees with the HQE result in Eq.~(\ref{Eq:incres2}). Therefore, local duality is unambiguously seen in the massless limit of quarks except the heavy decaying one, that is indeed an analogy of the Pauli interference \cite{Bigi:1999fi, Bigi:1999qe}. Moreover local duality in the heavy meson mixings is understood as an example of the ``exclusive'' duality \cite{Shifman:2000jv}, where one exclusive mode approximates the inclusive result. The heavy quark limit is unnecessary to derive duality in this case. Another point to mention is that the {\it twisted} sum over exclusive states in Eqs.~(\ref{Eq:exc1},  \ref{Eq:exc2}) asymptotically gives $\Gamma_{ij}^{(H,\mathrm{exc})}\to \mathrm{const.}$ while non-{\it twisted} sum, corresponding to non-leptonic decay, scales like $\Gamma_{ij}^{(H,\mathrm{nl})}\propto m_Q$ \cite{Grinstein:1997xk} so that whether the topology is twisted affects the asymptotic $m_Q$ dependence of the observables.
\section{Local duality for massive flavors}\label{Sec:4}\noindent
In reality, $s$ and $c$ quarks cannot be regarded as massless particles. Including these masses is crucial in the presence of the GIM mechanism, since otherwise the net observables vanish in the limit where a particular CKM product is neglected. To this end, in this section, we investigate local duality and its violation for those massive quarks by numerically solving the 't Hooft equation.
In Sec.~\ref{Sec:4.1}, duality in the contributions from individual flavors is discussed. Subsequently, the result for the GIM combination that appears in the observable is presented in Sec.~\ref{Sec:4.2}.
\par
In numerically solving the 't Hooft equation, standard methods adopted in the literature might be the Multhopp technique (see  Ref.~\cite{Multhopp} and also appendices in Refs.~\cite{Jaffe:1991ib, Grinstein:1997xk} for the detail), where the wave function is expanded by the trigonometric basis function. The integral equation is then regarded as an eigenvalue problem, yielding the asymptotically linear Regge trajectory of meson mass spectra. The normalization of the eigenvectors obtained is rescaled so as to satisfy $\int_{0}^{1}\D x [\phi(x)]^2=1$. It is often pointed out for the Multhopp technique, however, that the behaviors at the end points, $\phi_n(x) = x^{\beta_1}$ for $x\sim 0$ and $\phi_n(x) = (1-x)^{\beta_2}$ for $x\sim 1$ with $\beta_{1, 2}$ being $\pi \beta_{1, 2}\cot(\pi \beta_{1, 2})=\beta^2-m_{1, 2}^2$, are not straightforward to obtain. Then, the BSW-improved Multhopp method \cite{Brower:1978wm} is developed, rendering the behavior at the end points better controlled. In this work, we adopt the method in Ref.~\cite{Lebed:2000gm}, introducing the following expansion,
\bea
\phi_n(x)=\displaystyle\sum_{k=1}^{N}a_k^{(n)}\sin(k \theta),\quad \theta = \mathrm{arccos}(2x-1).\label{eq:N}
\eea
We then convert the 't Hooft equation into the eigenvalue problem with the recursive formula in Ref.~\cite{Lebed:2000gm}, where the accuracy nearby end points are improved by taking large $N$, and obtain $a_k^{(n)}$ and $M_n^2$. Nonetheless, the endpoint behaviors for $x\to0, 1$ are still given by square root, so that great care must be taken for the accuracy. As Q that forms the bound state of $Q\bar{q}$ gets heavier, the meson wave function at the vicinity of $x=1$ becomes rather singular. Excited states that are formed by light quark and anti-quark with large $n$, whose wave functions rapidly oscillate, also cause errors in the presence of the limited precision around the endpoints. In this work, we take $N$ in Eq.~(\ref{eq:N}) as 500 and solve the 't Hooft equation, and then truncate heavier $(500-N_{\mathrm{eff}})$ excited states, that do not follow the linear Regge trajectory, as well as eigenvectors. $N_{\rm eff}$ is varied to test the stability of the numerical results. Moreover, the numerical analysis requires the evaluation of the overlap integrals for the convolution of meson wave functions in Eqs.~(\ref{Eq:Ftri}-\ref{Eq:Ddou}), distinguished from the simpler one for semi-leptonic decays of heavy mesons in Ref.~\cite{Lebed:2000gm}. In order to guarantee the numerical stability of the result presented below, we neglect the triple overlap integral in Eq.~(\ref{Eq:Ftri}), that gives a contribution suppressed by at least $1/m_Q^2$ \cite{Bigi:1999fi} to the decay amplitudes, relative to the leading terms in Eqs.~(\ref{Eq:Cdou}, \ref{Eq:Ddou}). For this case, the stability under the variation of $N_{\rm eff}$ is verified. Hereafter we fix $N_{\rm eff}=300$. Further improvement in the numerical results entails technical tasks, including the accurate calculation of endpoint behaviors, as well as the precise evaluation of the convolution integral, which are beyond our current scope, while the exclusive results presented below capture the leading behaviors in the $1/m_Q$ expansion. As was obtained in Sec.~\ref{Sec:3.2}, the HQE result includes the term proportional to $(c_{\rm V}^4-c_{\rm A}^4)$ in addition to one multiplied by $(c_{\rm V}^2-c_{\rm A}^2)^2$, where the former is not included in the exclusive results in Eqs.~(\ref{Eq:exc1}, \ref{Eq:exc2}) with Eq.~(\ref{Eq:topo}). For comparing inclusive and sum of exclusive result in a consistent manner, we take only the terms proportional to $(c_{\rm V}^2-c_{\rm A}^2)^2$ in Eqs.~(\ref{Eq:CA}, \ref{Eq:CP}) in what follows.
\par
Before proceeding to results, further remarks are addressed: 
\begin{itemize}
\item For the heavy quark decays, spikes of the rate emerge \cite{Grinstein:1997xk, Grinstein:1998gc} when the heavy quark mass gets larger than threshold values for $M_{H}=M_k+M_m$ due to the hadronic phase space unlike the case in $3+1$ dimensions. In order to quantify violation of local duality, the middle point between $i$-th and $(i+1)$-th thresholds should be discussed \cite{Bigi:1998kc}. For the width difference in the heavy meson mixings, the analogous spikes appears for massive final states, as well as decays. The numerical results presented below are based on discrete points for heavy quark mass that are not (exactly) at the thresholds to avoid obvious singularities in Eqs.~(\ref{Eq:exc1}, \ref{Eq:exc2}).
\item In principle, bare masses and a bare coupling for $d=2$ have no intrinsic relations to ones for $d=4$. For an illustrative reason, we take reference values of bare masses for $d=2$ as central values from PDG \cite{Zyla:2020zbs} as
$\overline{m}_s(2~\mathrm{GeV})=93~\mathrm{MeV}$,
$\overline{m}_c(\overline{m}_c)=1.280~\mathrm{GeV}$,
$m_c^{\mathrm{pole}}=1.67~\mathrm{GeV}$,
$\overline{m}_b(\overline{m}_b)=4.18~\mathrm{GeV}$,
$m_b^{1S}=4.65~\mathrm{GeV}$,
$m_b^{\mathrm{pole}}=4.78~\mathrm{GeV}$. In the calculation of the $B^0_s-\bar{B}^0_s$ mixing, the bare mass of strange quark is fixed by the $\overline{\rm MS}$ mass of the strange quark mass at the scale of bottom quark mass evaluated by the renormalization group evoluation \cite{Chetyrkin:2000yt} for $d=4$. The bare masses for $u$ and $d$ quarks are fixed to zero in what follows. As for the bare coupling, we adopt an ansatz, $\beta=340~\mathrm{MeV}$, that is obtained in such a way that the string tension of QCD$_4$ is fitted \cite{Burkardt:2000uu, Jia:2017uul} by $(\pi /2)\beta^2=0.18~\mathrm{GeV}^2$.
\end{itemize}
\subsection{Numerical result for individual flavors}
\label{Sec:4.1}
\begin{figure}[H]
\vspace{-22mm}
\begin{center}
\subfigure[]{\includegraphics[width=0.95\columnwidth]{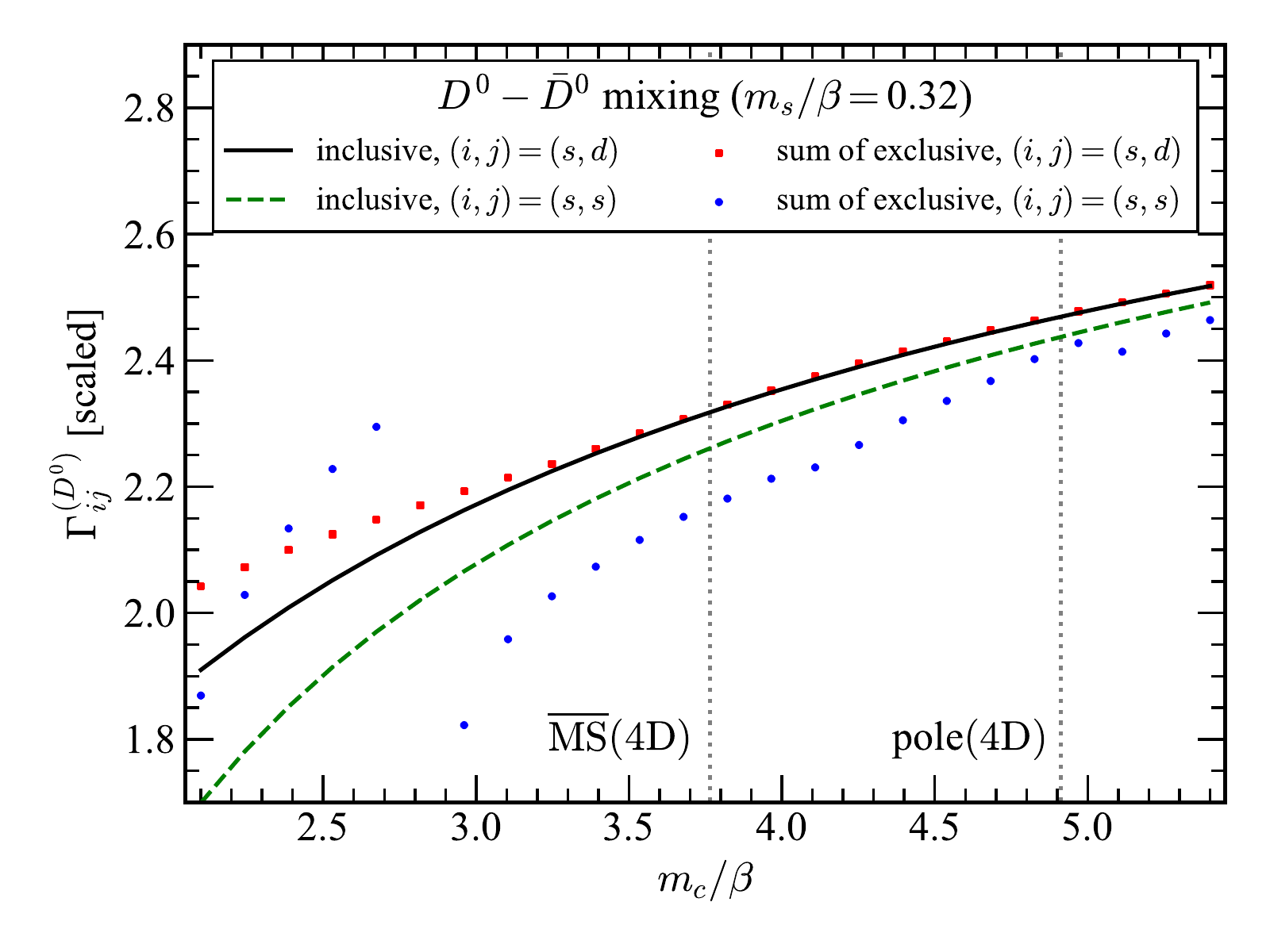}}\\
\vspace{0mm}
\subfigure[]{\includegraphics[width=0.95\columnwidth]{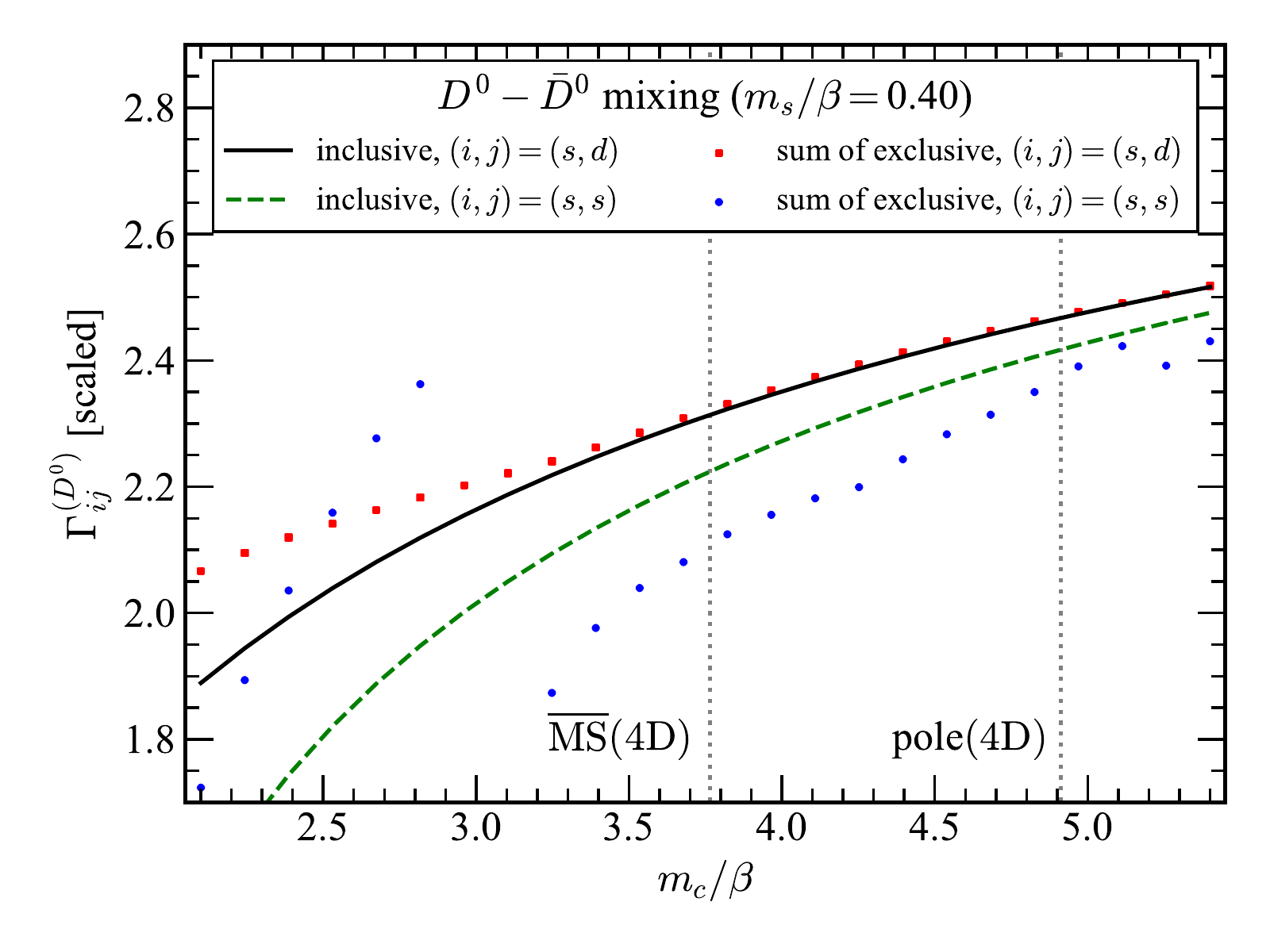}}
\end{center}
\vspace{-5mm}
\caption{Comparison between inclusive and sum of exclusive width differences for the $D^0-\bar{D}^0$ mixing: (a) $m_s/\beta=0.32$, (b) $m_s/\beta=0.40$. The black solid (green dashed) line represents the inclusive result for one (two) massive intermediate state(s) while the red square (blue point) stands for the sum of exclusive results from one (two) massive quark(s) in the final state. The dotted vertical lines correspond to reference values of the masses in four-dimensions. The vertical axis is normalized by $4G_F^2(c_{\rm V}^2-c_{\rm A}^2)^2\beta N_c/\pi$.}
\label{Fig:3}
\end{figure}

\begin{figure}[H]
\vspace{-22mm}
\begin{center}
\subfigure[]{\includegraphics[width=0.95\columnwidth]{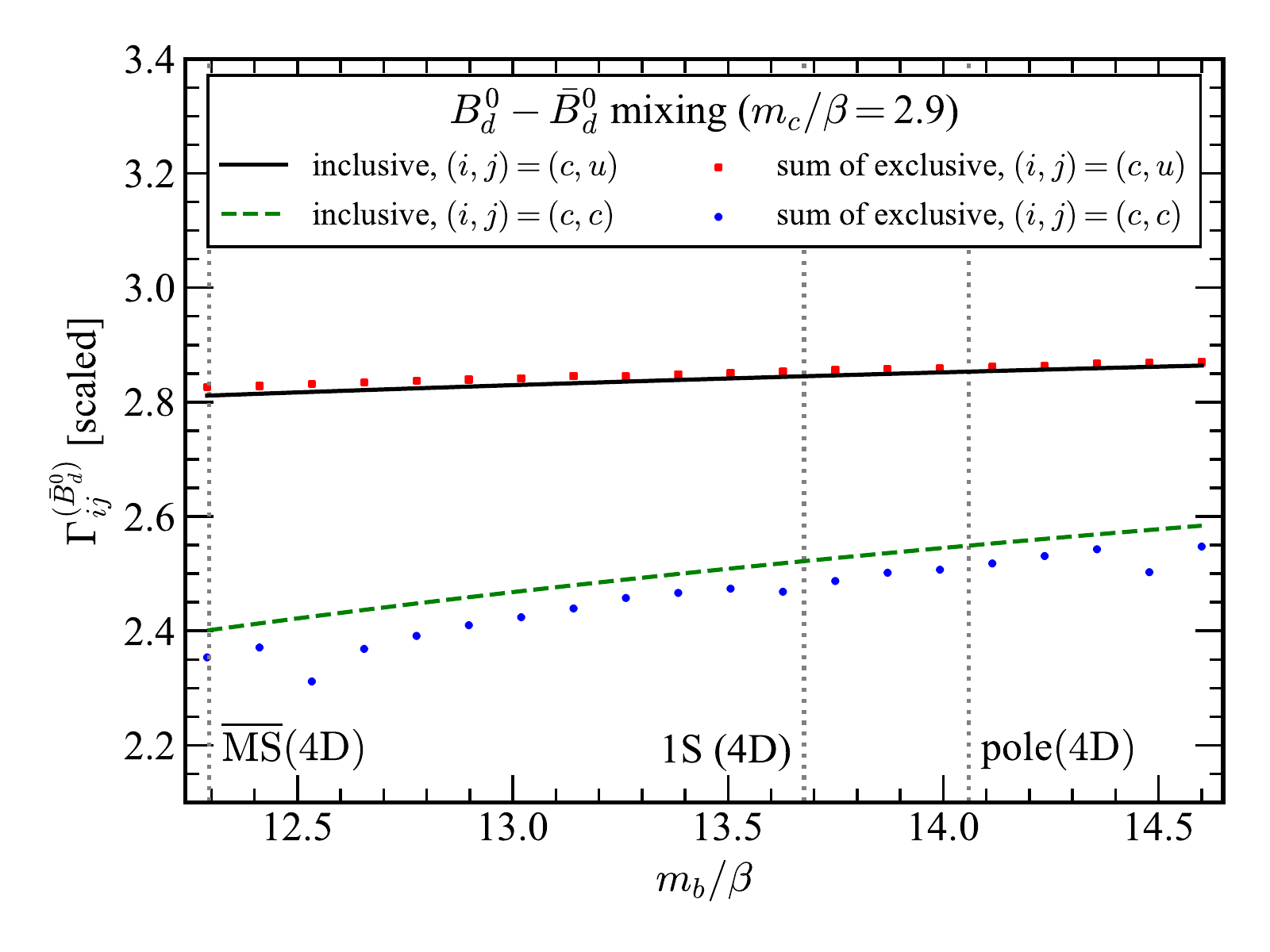}}\\
\vspace{0mm}
\subfigure[]{\includegraphics[width=0.95\columnwidth]{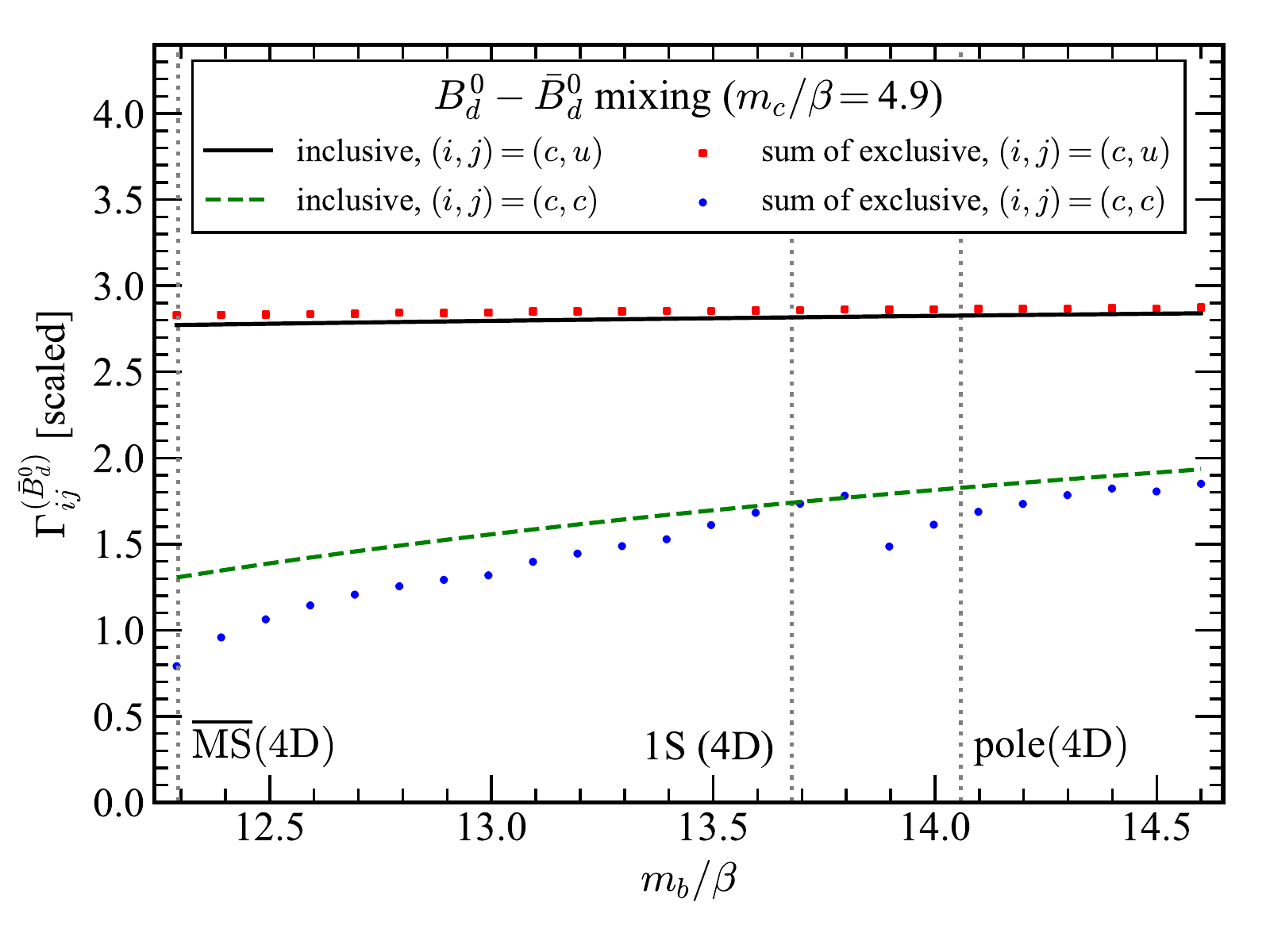}}
\end{center}
\vspace{-5mm}
\caption{Same as Fig.~\ref{Fig:3} for $B^0_d-\bar{B}^0_d$ mixing: (a) $m_c/\beta=2.9$, (b) $m_c/\beta=4.9$.}
\label{Fig:4}
\end{figure}

\begin{figure}[H]
\vspace{-22mm}
\begin{center}
\subfigure[]{\includegraphics[width=0.95\columnwidth]{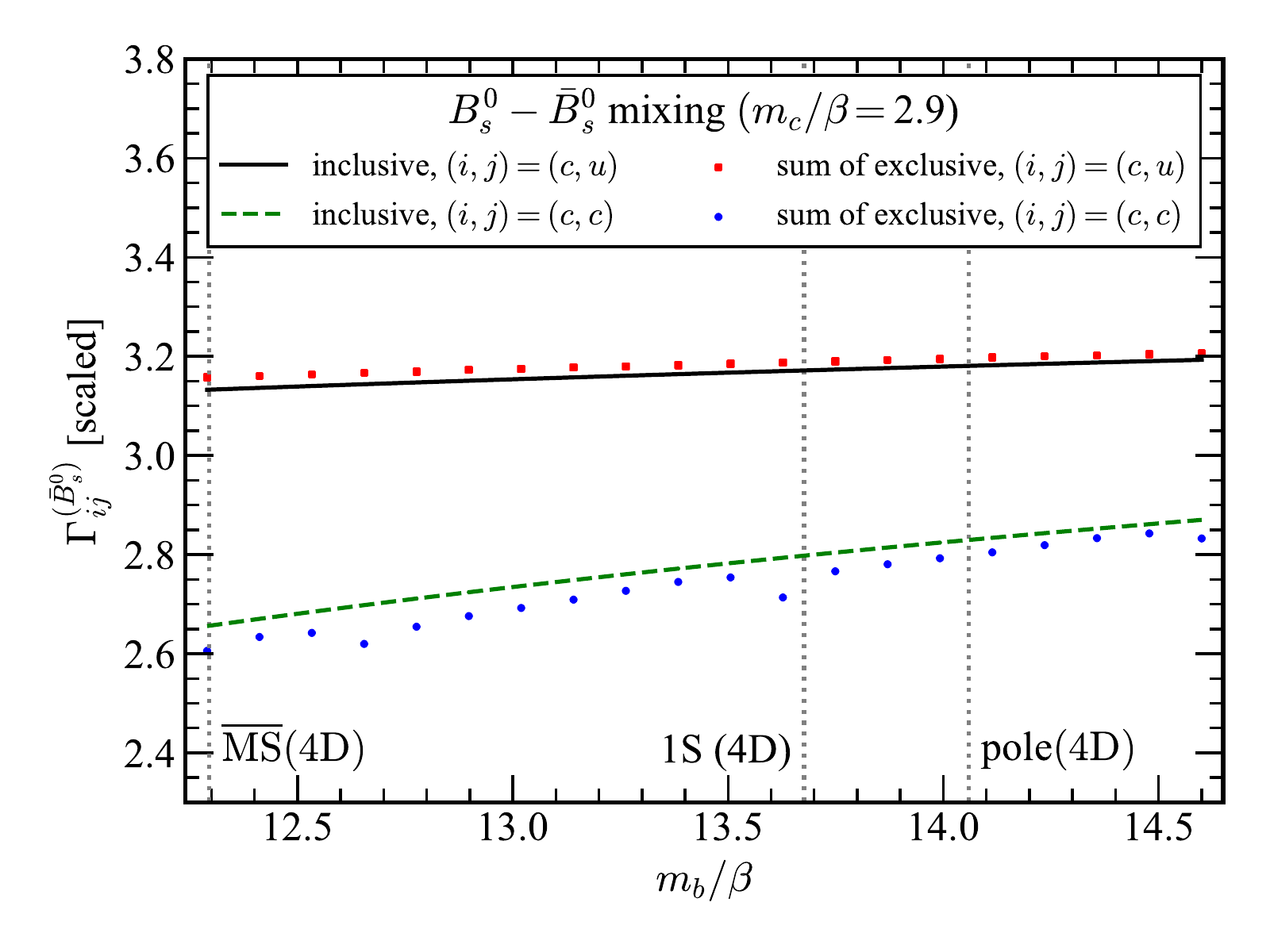}}\\
\subfigure[]{\includegraphics[width=0.95\columnwidth]{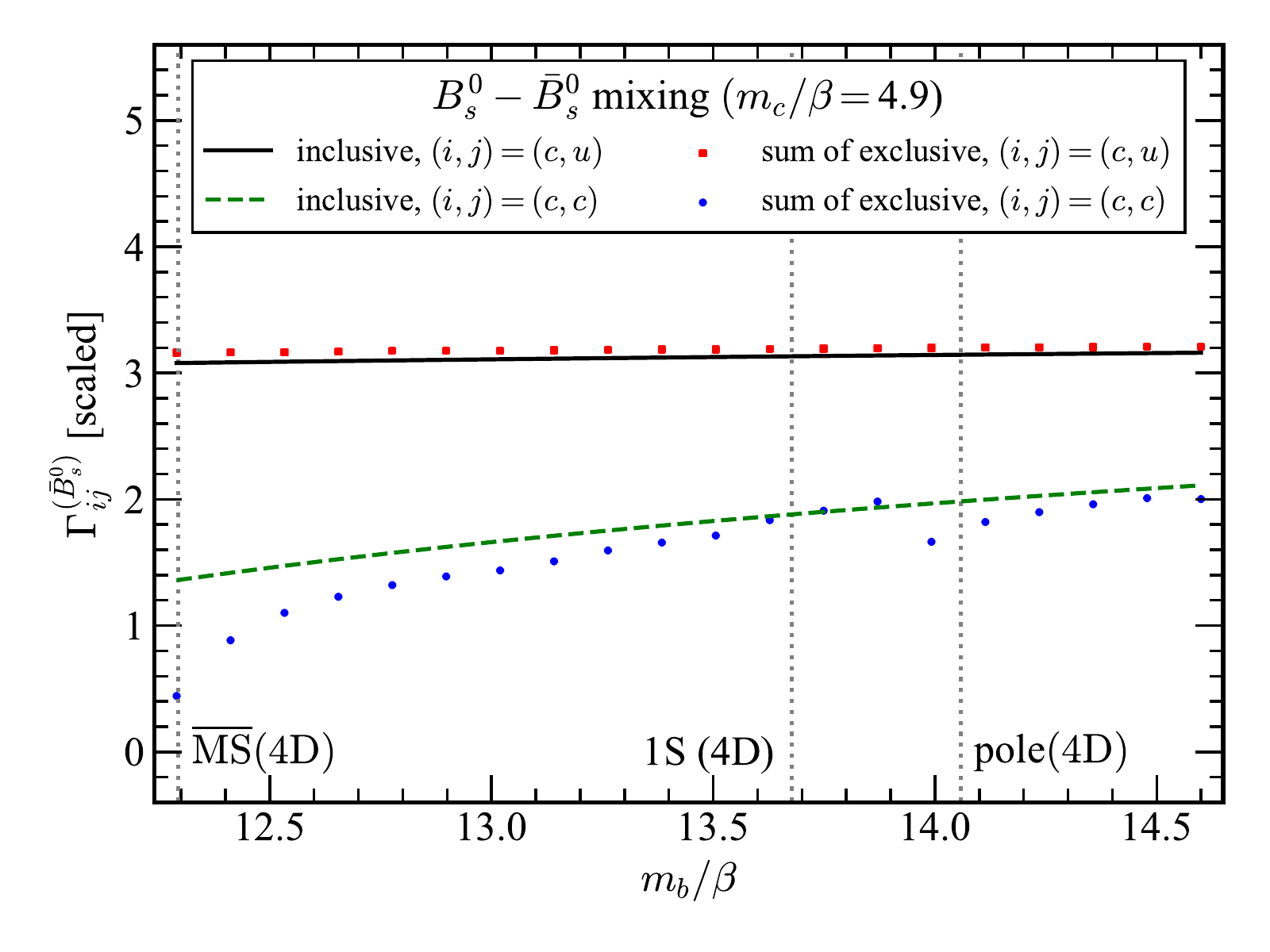}}
\end{center}
\vspace{-5mm}
\caption{Same as Fig.~\ref{Fig:3} for $B^0_s-\bar{B}^0_s$ mixing: (a) $m_c/\beta=2.9$, (b) $m_c/\beta=4.9$.}
\label{Fig:5}
\end{figure}

\noindent
Both inclusive and sum of exclusive width differences for the $D^0-\bar{D}^0$, $B^0_d-\bar{B}^0_d$ and $B^0_s-\bar{B}^0_s$ mixings are exhibited in Figs.~\ref{Fig:3}-\ref{Fig:5}. The value of $\beta$ affects only the normalization of the vertical axes of the plots and also the locations of vertical lines showing quark mass in four-dimensions. Figure~\ref{Fig:3}a (\ref{Fig:4}a) is based on $m_s/\beta =0.32$  ($m_c/\beta =2.9$) corresponding to the $\overline{\rm MS}$ mass at the scale of charm (bottom) quark while Fig.~\ref{Fig:3}b (\ref{Fig:4}b) shows the result for $m_s/\beta= 0.40$ ($m_c/\beta=4.9$ corresponding to the pole mass). In each panel, two types of the width difference including one or two massive flavors, {\it i.e.,} $sd$ and $ss$ intermediate states for the $D^0-\bar{D}^0$ mixing and $cu$ and $cc$ intermediate states for the $B^0_q-\bar{B}^0_q$ $(q=d, s)$, are shown. Results similar to Fig.~\ref{Fig:4} except that the $B_d^0-\bar{B}_d^0$ mixing is replaced by the $B_s^0-\bar{B}_s^0$ mixing are exhibited in Fig.~\ref{Fig:5}. In addition to the results plotted in Figs.~\ref{Fig:3}-\ref{Fig:5}, there are also $\Gamma_{dd}^{(D^0, \alpha)}, \Gamma_{uu}^{(\bar{B}^0_d, \alpha)}$ and $\Gamma_{uu}^{(\bar{B}^0_s, \alpha)}$ $(\alpha =\mathrm{exc}, \mathrm{inc})$, that are not presented in the figures. Since those cases include the massless intermediate quarks, the numerical results should be consistent with the analytical results in Sec.~\ref{Sec:3.3}. Indeed, the reasonable agreement between inclusive and sum of exclusive width differences is numerically confirmed for all of the three cases including $\Gamma_{uu}^{(B^0_s, \mathrm{exc})}$ based on massive intermediate kaons, in which case the analytical discussion in the massless limit is not applied.
\par
One can find that for the $\Gamma_{ss}^{(D^0, \mathrm{exc})}, \Gamma_{cc}^{(\bar{B}^0_d, \mathrm{exc})}$ and $\Gamma_{cc}^{(\bar{B}^0_s, \mathrm{exc})}$ in Figs.~\ref{Fig:3}-\ref{Fig:5}, the spikes for width differences when the heavy quark mass gets larger than the threshold values are shown obviously. These are to be contrasted with the results for $\Gamma_{sd}^{(D^0, \mathrm{exc})}, \Gamma_{cu}^{(\bar{B}^0_d, \mathrm{exc})}$ and $\Gamma_{cu}^{(\bar{B}^0_s, \mathrm{exc})}$. The absence for the obvious threshold singularities for the latter three cases can be understood analytically as follows: we take $\Gamma_{sd}^{(D^0, \mathrm{exc})}$ as an example while the similar discussion is applied for $\Gamma_{cu}^{(\bar{B}^0_d, \mathrm{exc})}$. Due to the vanishing properties of decay constants for the excited states of pions, we find that the sum over pion states in Eq.~(\ref{Eq:exc1}) is reduced only to the ground state, as was discussed in Sec.~\ref{Sec:3.3}, so that,
\bea
\Gamma^{(D^0,\:\mathrm{exc})}_{ds}&=&\displaystyle\sum_{m}(-1)^{m}\frac{T_{(c\bar{u})(d, s)}^{(0, m)} T_{(c\bar{u})(s, d)}^{(m, 0)\:*}}{2M_{D}(M_D^2-M_m^2)}.\label{Eq:exam}
\eea
By recalling that in the massless limit of $u$ and $d$ quarks, the only surviving contribution in $T_{(c\bar{u})(d, s)}^{(0, m)}$ arises from the contact interaction term in Eq.~(\ref{Eq:Cdou}), one finds,
$T_{(c\bar{u})(d, s)}^{(0, m)}\propto q^2(1-\omega)/\omega=(M_{D}^2-M_m^2)$ in the limit of $q^2\to 0$ together with $\omega\to 0$. Substituting this relation into Eq.~(\ref{Eq:exam}), we find that the phase space singularities for each threshold of $m$ cancel out with the decay amplitude of $D^0\to \pi^{+(0)} K^{-(m)}$.
\par
Moreover, for $\Gamma_{sd}^{(D^0, \mathrm{exc})}, \Gamma_{cu}^{(\bar{B}^0_d, \mathrm{exc})}$ and $\Gamma_{cu}^{(\bar{B}^0_s, \mathrm{exc})}$, when the heavy quark mass is large, the agreement between inclusive and sum of exclusive width differences is better than $\Gamma_{ss}^{(D^0, \mathrm{exc})}, \Gamma_{cc}^{(\bar{B}^0_d, \mathrm{exc})}$ and $\Gamma_{cc}^{(\bar{B}^0_s, \mathrm{exc})}$ although the analytical understanding for this remains unclear (the coincidence is slowly improved for $\Gamma_{cu}^{(\bar{B}^0_d, \alpha)}$ and $\Gamma_{cu}^{(\bar{B}^0_s, \alpha)}$ in the plotted domains of Figs.~\ref{Fig:4}-\ref{Fig:5}). Consequently, it is expected that patterns of flavor symmetry breaking in (GIM, 1) given in Eqs.~(\ref{Eq:GIM1}, \ref{Eq:GIM3}) is rather different from (GIM, 2) in Eqs.~(\ref{Eq:GIM2}, \ref{Eq:GIM4}) in the currently considered case. 

\subsection{Numerical result in the presence of the GIM mechanism}
\label{Sec:4.2}\noindent
We would like to remind the reader that the inclusive width difference, discussed in Sec.~\ref{Sec:3.2}, has quite different function forms, depending on whether (1) 4D-like phase space term in Eq.~(\ref{Eq:phasespace0}) is only considered or (2) the 2D-specific terms in Eqs.~(\ref{Eq:phasespaceG}, \ref{Eq:phasespaceH}) are additionally included. For the former, the GIM 1 for the $D^0-\bar{D}^0$ mixing defined in Eq.~(\ref{Eq:GIM1}) behaves like $(m_s/m_c)^4$ while it is $(m_s/m_c)^2$ for the latter in the large $m_c$ limit, due to Eq.~(\ref{Eq:expD1}) and Eq.~(\ref{Eq:expD3}), respectively, meaning that the former is more suppressed. The similar discussion is applied for the $B_q^0-\bar{B}_q^0$ mixing by replacing $m_c\to m_b$ and $m_s\to m_c$. Thus, the order of the magnitude of $|\Gamma^{\rm (exc)}/\Gamma^{\rm (inc)}|$ strongly depends on whether (1) or (2) is adopted for the inclusive side. Below, we present the results based on both (1) and (2).
\par
\begin{figure}[H]
\vspace{-22mm}
\begin{center}
\subfigure[]{\includegraphics[width=0.95\columnwidth]{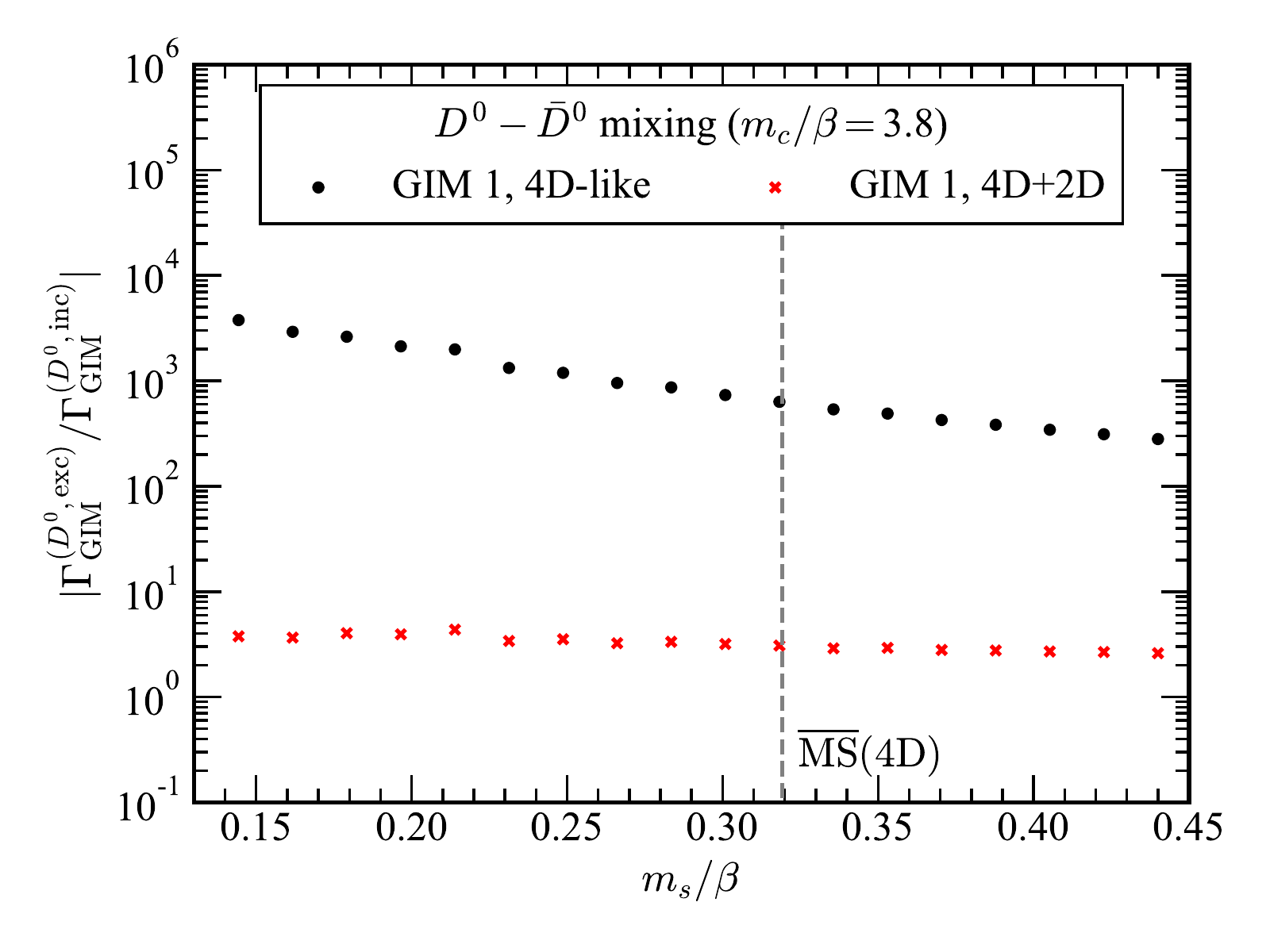}}\\
\subfigure[]{\includegraphics[width=0.95\columnwidth]{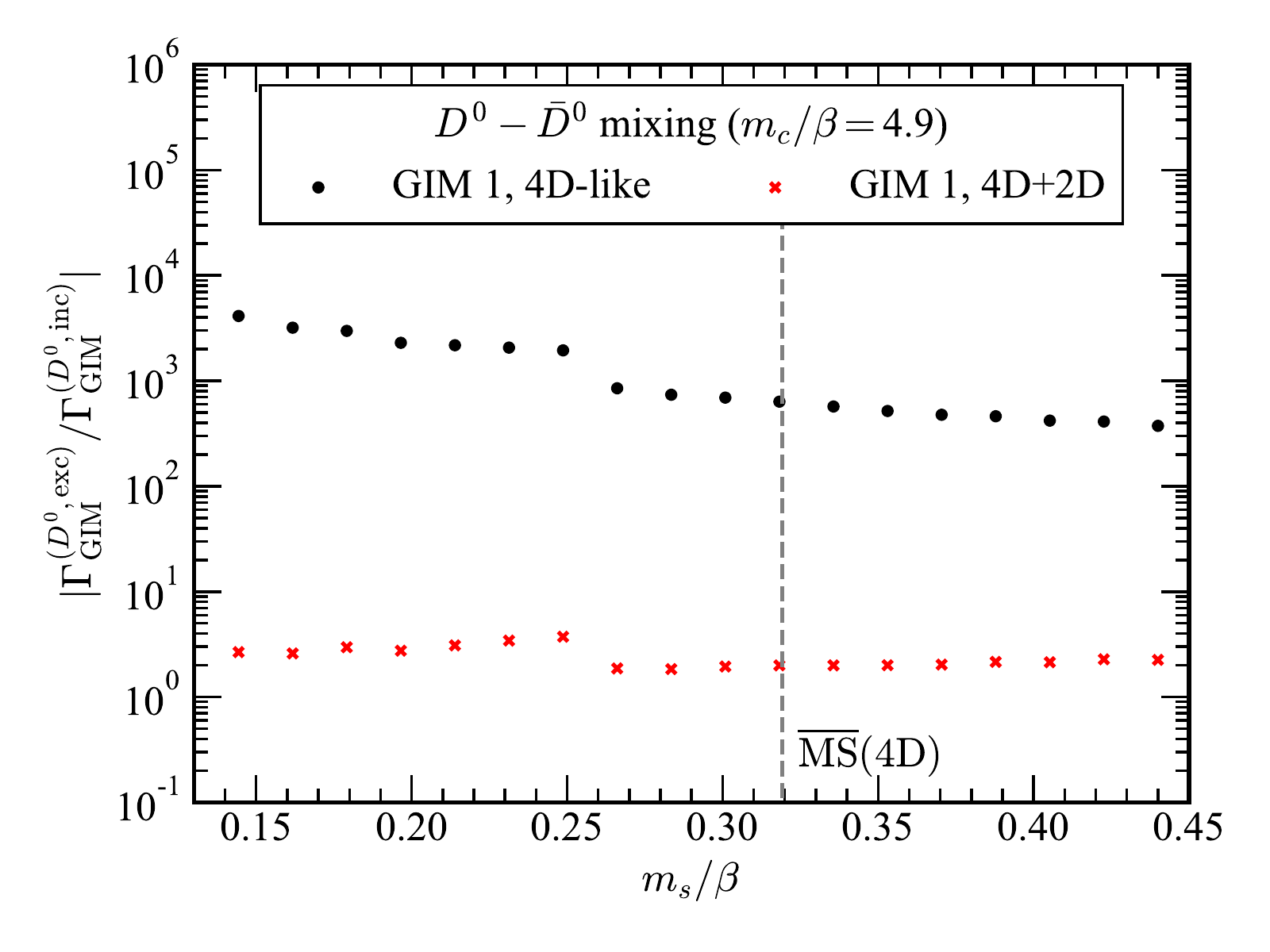}}
\end{center}
\vspace{-5mm}
\caption{Absolute values for the ratios of the exclusive GIM 1 in Eqs.~(\ref{Eq:GIM1}, \ref{Eq:GIM3}) to the inclusive one for the $D^0-\bar{D}^0$ mixing: (a) $m_c/\beta=3.8$, (b) $m_c/\beta=4.9$. The black dot represents the case where only 4D-like phase space term in Eq.~(\ref{Eq:phasespace0}) is considered in the inclusive calculation while the red cross corresponds to the case where both 4D-like and 2D-specific terms in Eqs.~(\ref{Eq:phasespace0}-\ref{Eq:phasespaceH}) except for Eq.~(\ref{Eq:phasespace}) are included.}
\label{Fig:6}
\end{figure}
\begin{figure}[H]
\vspace{-22mm}
\begin{center}
\subfigure[]{\includegraphics[width=0.95\columnwidth]{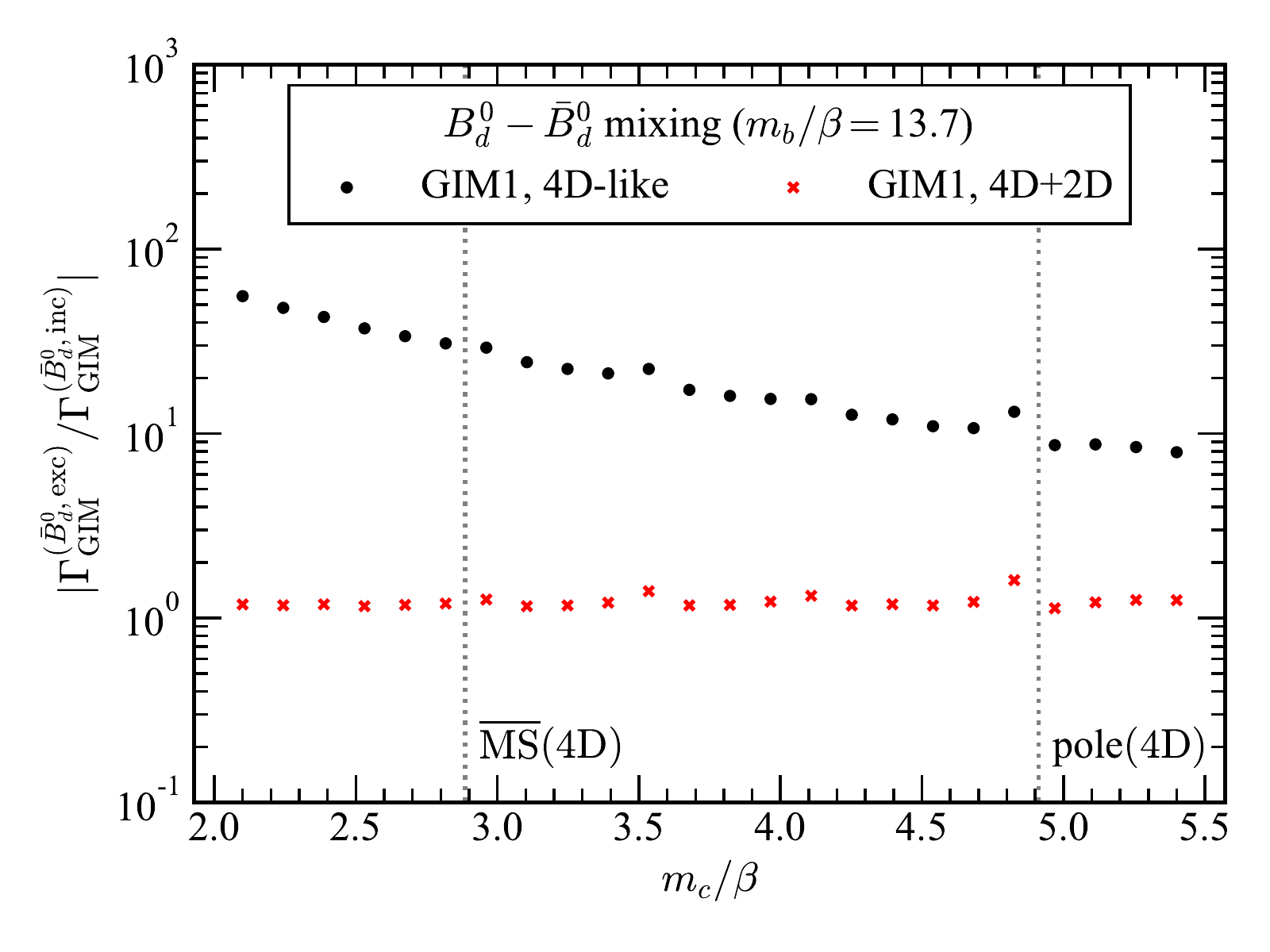}}\\
\subfigure[]{\includegraphics[width=0.95\columnwidth]{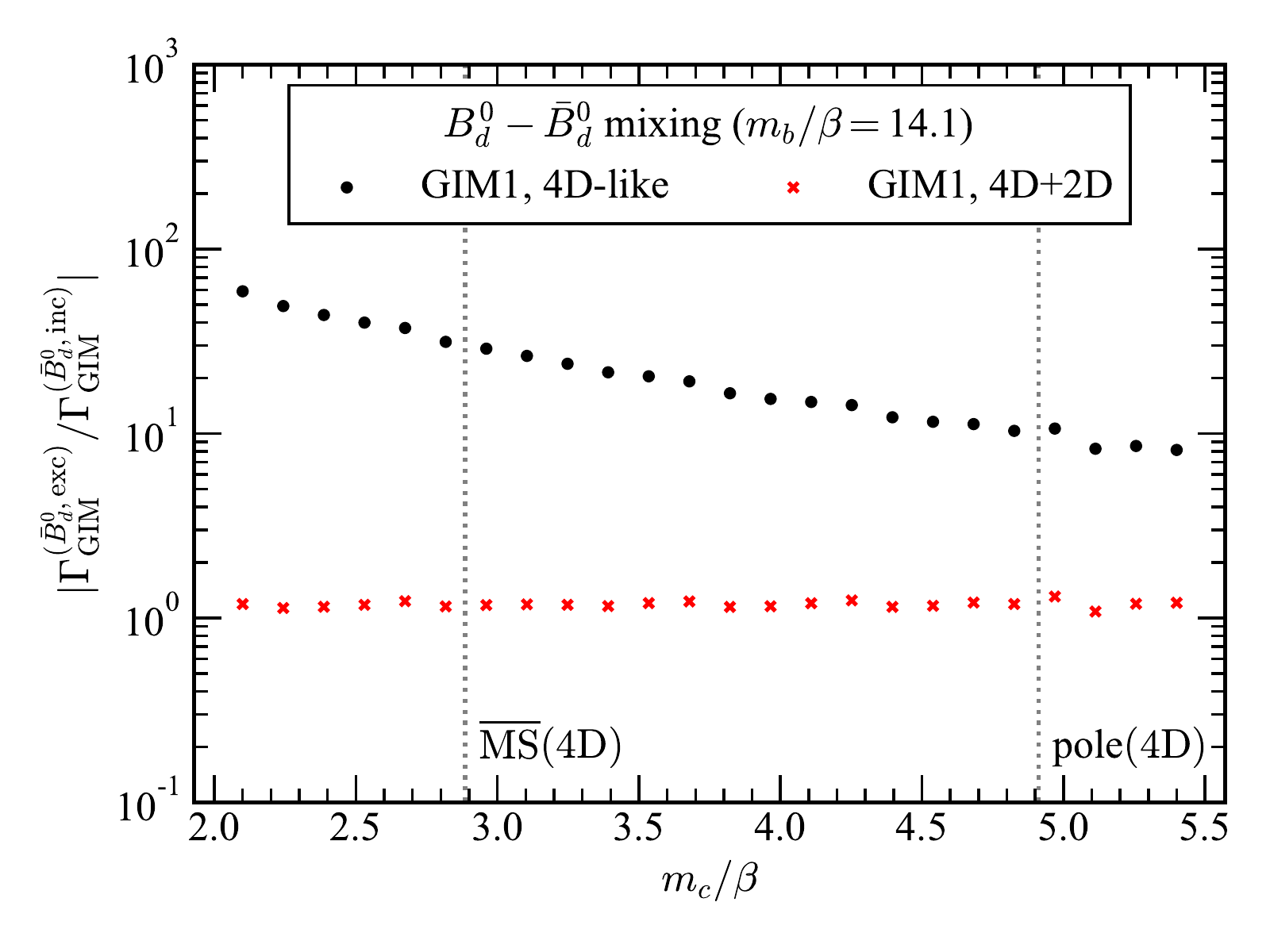}}
\end{center}
\vspace{-5mm}
\caption{Same as Fig.~\ref{Fig:6} for the $B^0_d-\bar{B}^0_d$ mixing: (a) $m_b/\beta=13.7$, (b) $m_b/\beta=14.1$.}
\label{Fig:7}
\end{figure}
\begin{figure}[H]
\vspace{-22mm}
\begin{center}
\subfigure[]{\includegraphics[width=0.95\columnwidth]{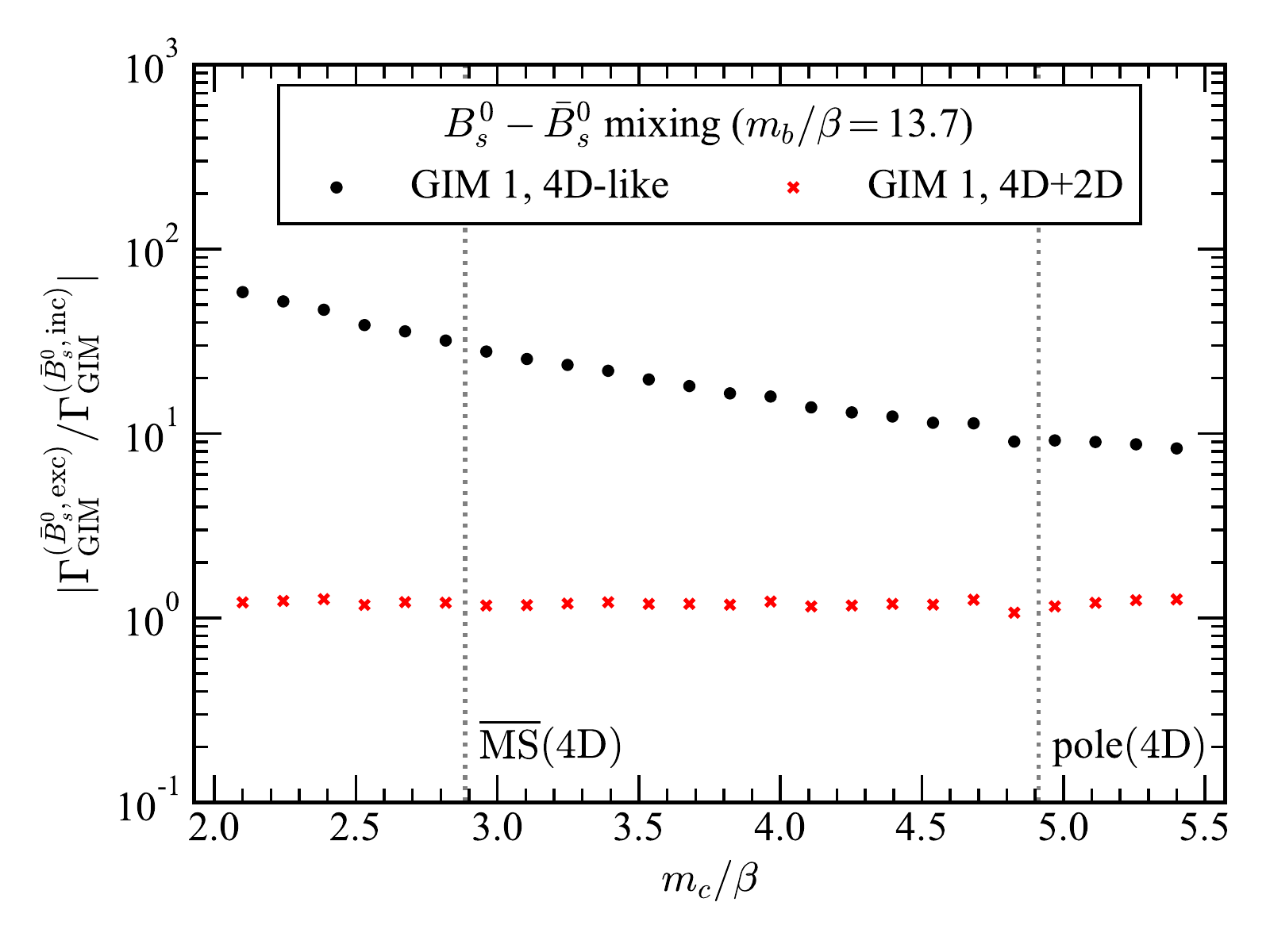}}\\
\subfigure[]{\includegraphics[width=0.95\columnwidth]{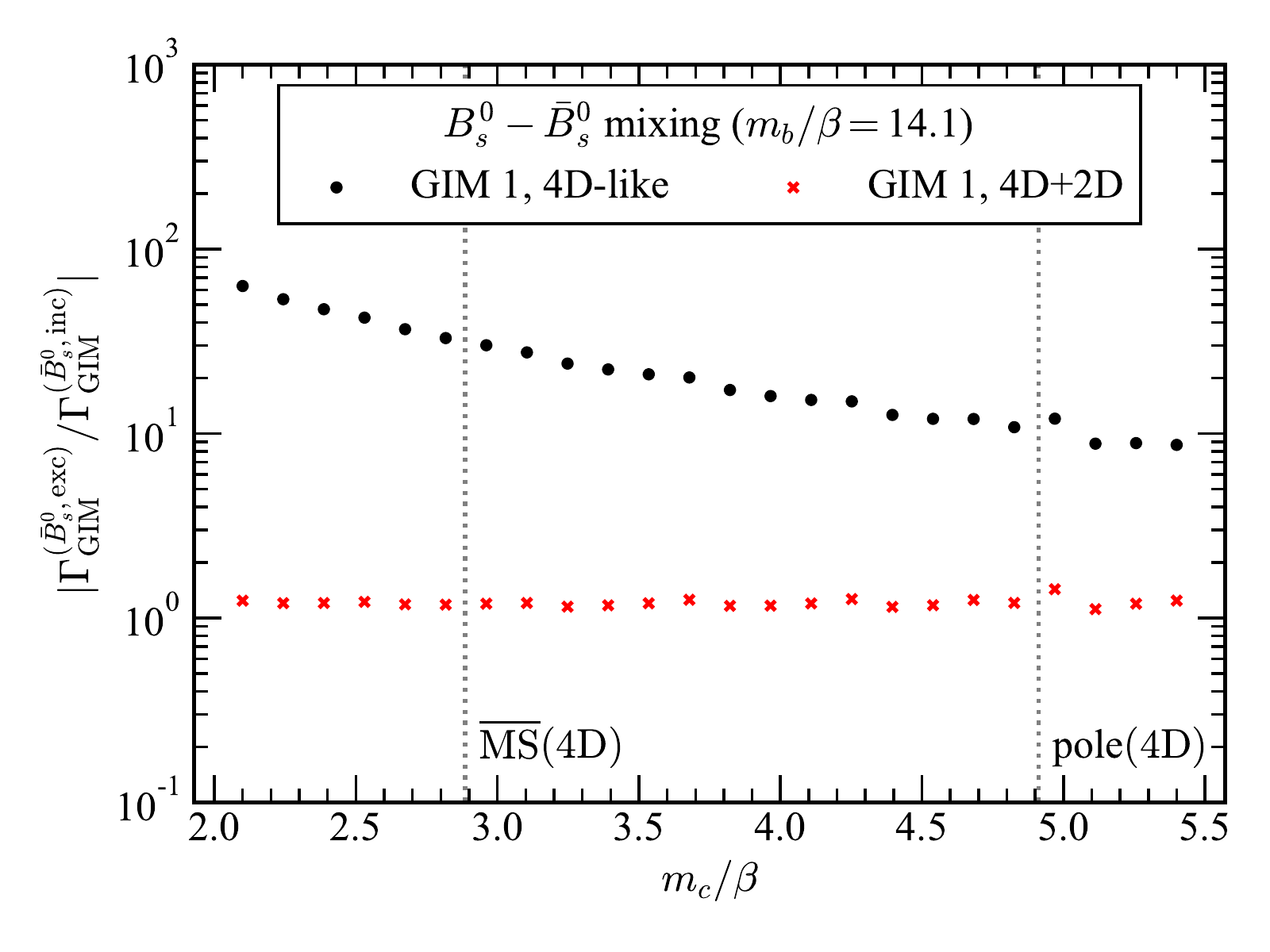}}
\end{center}
\vspace{-5mm}
\caption{Same as Fig.~\ref{Fig:6} for the $B^0_s-\bar{B}^0_s$ mixing: (a) $m_b/\beta=13.7$, (b) $m_b/\beta=14.1$.}
\label{Fig:8}
\end{figure}

In Figs.~\ref{Fig:6}-\ref{Fig:8}, absolute values for the ratio of exclusive the GIM 1 combination to inclusive one defined both in Eqs.~(\ref{Eq:GIM1}, \ref{Eq:GIM3}) are given for the three meson mixings. The two panels in each figure are associated with different choices of bare masses for the external quarks. The $\overline{\rm MS}$ masses shown as reference values are evaluated at the scale of the external heavy quark mass for $d=4$. For the $D^0-\bar{D}^0$ mixing, the enhancement of the exclusive result is larger than $10^3$ for $m_s/\beta<0.25$ when the inclusive rate includes only the 4D-like phase space term, $F_{ij}^{\rm (th)}$ in Eq.~(\ref{Eq:phasespace0}). As for the $B^0_q-\bar{B}^0_q$ $(q=d, s)$ mixing, a similar enhancement is observed when only the 4D-like phase space term is included, although the enhancement for the $B_q^0-\bar{B}_q^0$ mixing is not as strong as the $D^0-\bar{D}^0$ mixing. The pattern for the $B^0_d-\bar{B}^0_d$ mixng in Fig.~\ref{Fig:7} is similar to that of $B^0_s-\bar{B}^0_s$ in Fig.~\ref{Fig:8}. Except that the plotted ratios undergo some jumps when the external quark mass crosses the hadronic thresholds, the results are given by regular curves in all of Figs.~\ref{Fig:6}-\ref{Fig:8}. The dumping behaviors of the results in Figs.~\ref{Fig:6}-\ref{Fig:8} based on only the 4D-like phase space term as the external quark mass is enlarged indicate that the sum of the exclusive width difference is scaled as $\Gamma_{(\rm GIM, 1)}^{(D^0, \mathrm{exc})} \propto m^n_s$ and $\Gamma_{(\rm GIM, 1)}^{(\bar{B}^0_q, \mathrm{exc})} \propto m^n_c$ with $n<4$ since the 4D-like inclusive width difference behaves like $n=4$ as shown in Eq.~(\ref{Eq:expD1}). It should be noted that for the $D^0-\bar{D}^0$ mixing the quantity plotted in Fig.~\ref{Fig:6} is of direct relevance in phenomenology, while this is not the case for the $B^0_q-\bar{B}^0_q$ $(q=d, s)$ mixing, as was discussed in Sec.~\ref{Sec:2.2}. The numerical stabilities under the variation of $N_{\rm eff}$ are confirmed for what are plotted in Figs.~\ref{Fig:6}-\ref{Fig:8}, especially $m_s/\beta>0.14$ in Fig.~\ref{Fig:6}.
\par
\begin{figure}[H]
\vspace{-22mm}
\begin{center}
\subfigure[]{\includegraphics[width=0.95\columnwidth]{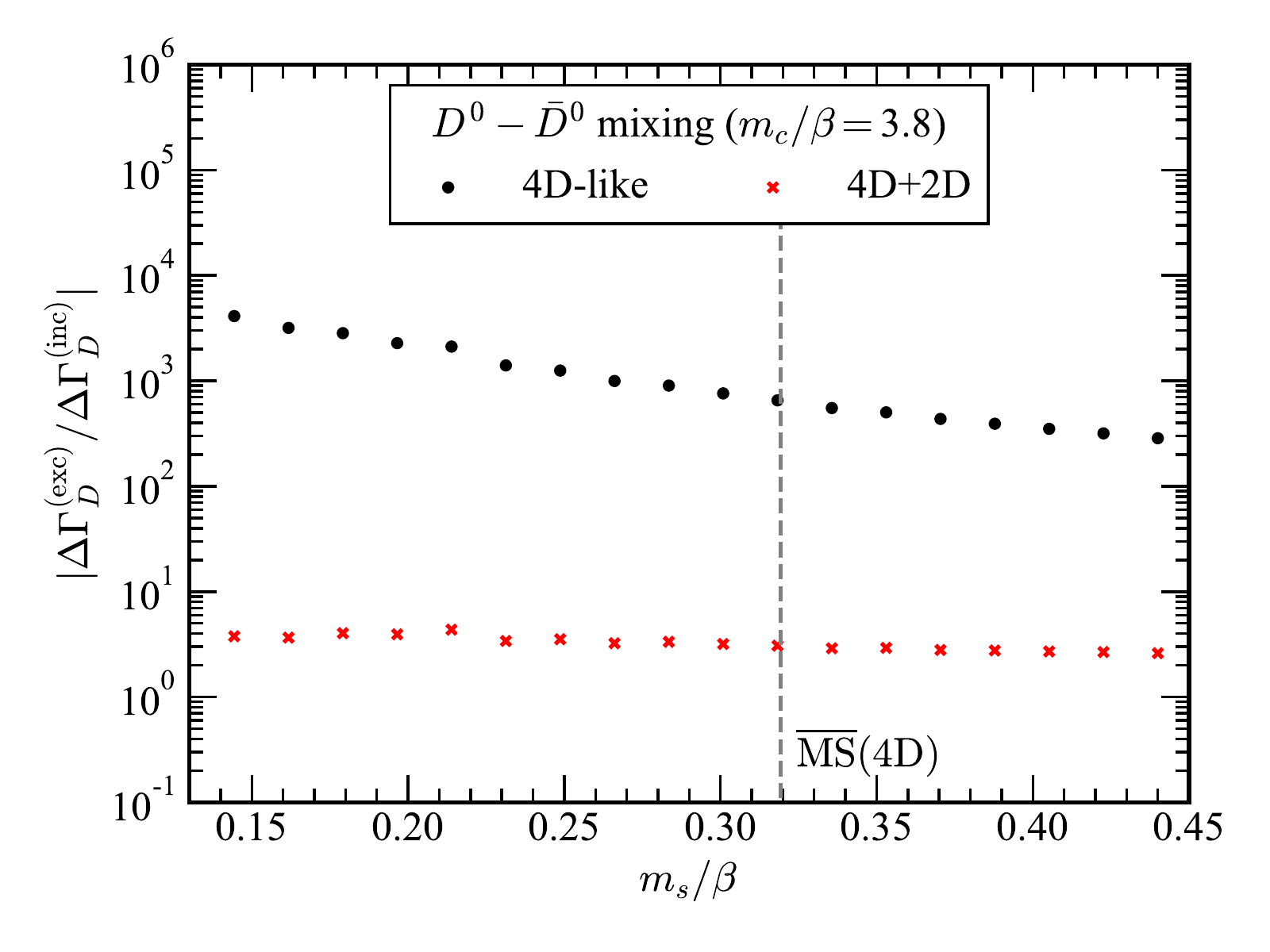}}\\
\subfigure[]{\includegraphics[width=0.95\columnwidth]{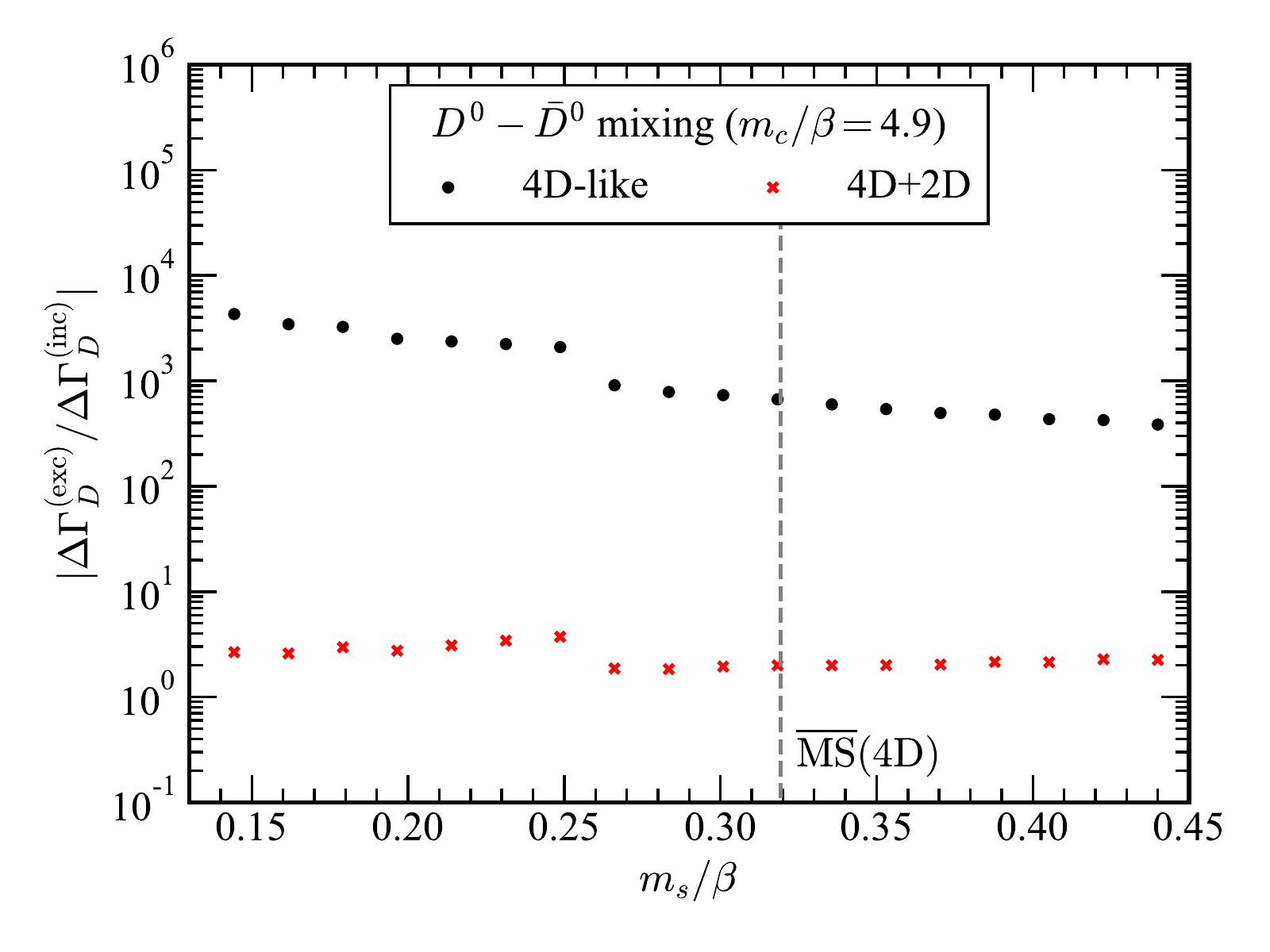}}
\end{center}
\vspace{-5mm}
\caption{Same as Fig.~\ref{Fig:6} for the observable width difference in the $D^0-\bar{D}^0$ mixing: (a) $m_c/\beta=3.8$, (b) $m_c/\beta=4.9$.}
\label{Fig:9}
\end{figure}
\begin{figure}[H]
\vspace{-22mm}
\begin{center}
\subfigure[]{\includegraphics[width=0.95\columnwidth]{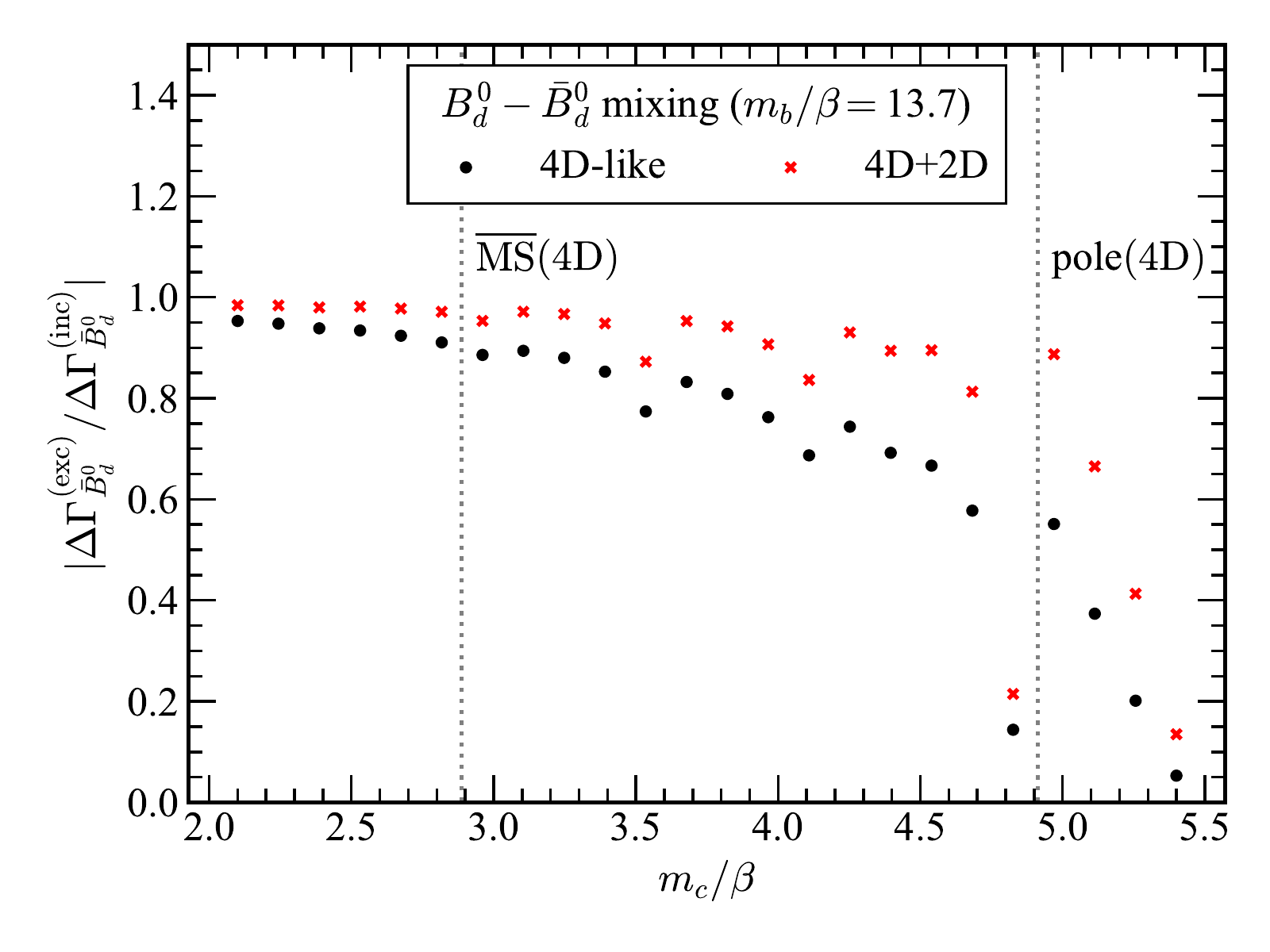}}\\
\subfigure[]{\includegraphics[width=0.95\columnwidth]{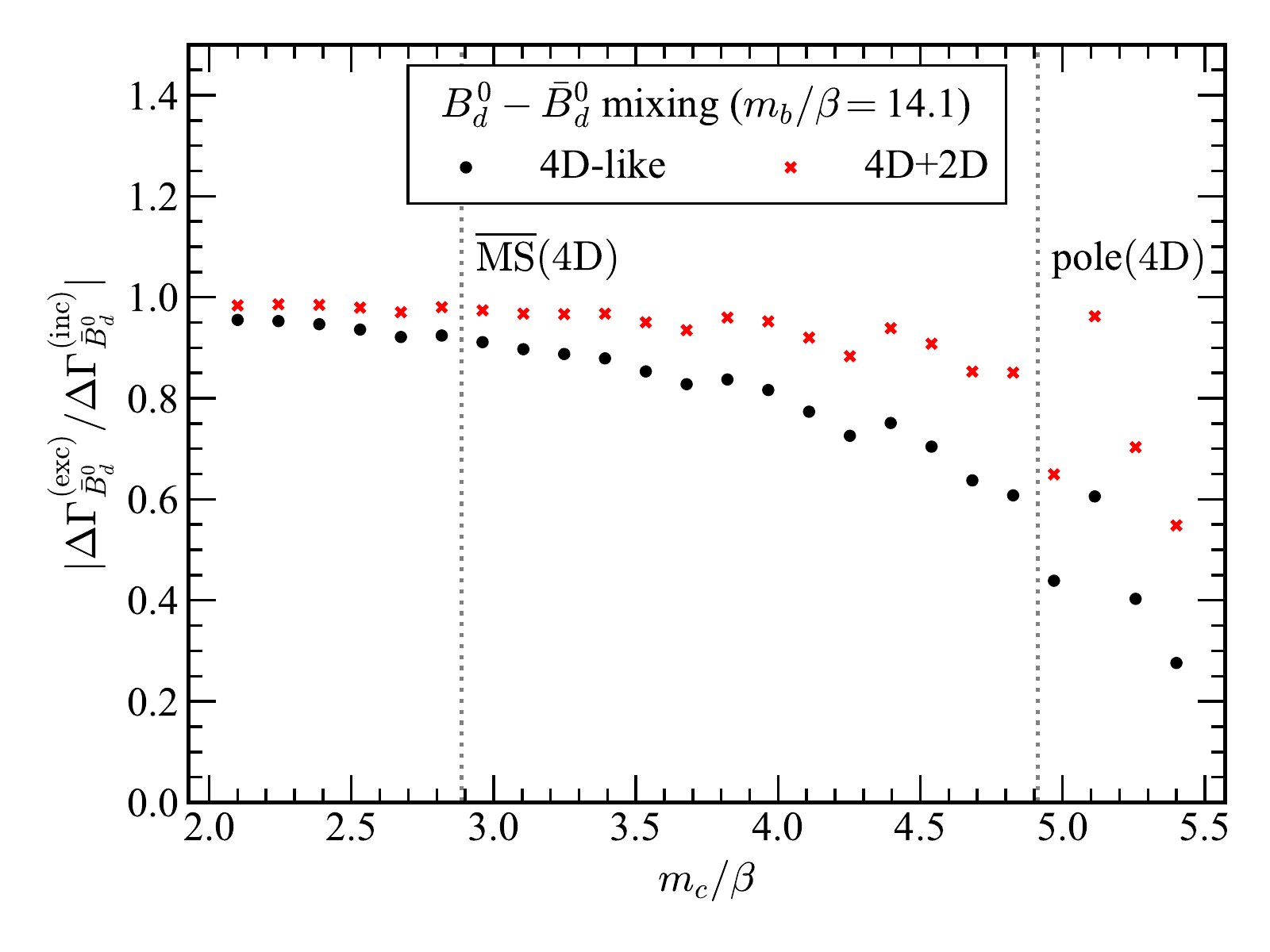}}
\end{center}
\vspace{-5mm}
\caption{Same as Fig.~\ref{Fig:9} for the $B^0_d-\bar{B}^0_d$ mixing: (a) $m_b/\beta=13.7$, (b) $m_b/\beta=14.1$.}
\label{Fig:10}
\end{figure}
\begin{figure}[H]
\vspace{-22mm}
\begin{center}
\subfigure[]{\includegraphics[width=0.95\columnwidth]{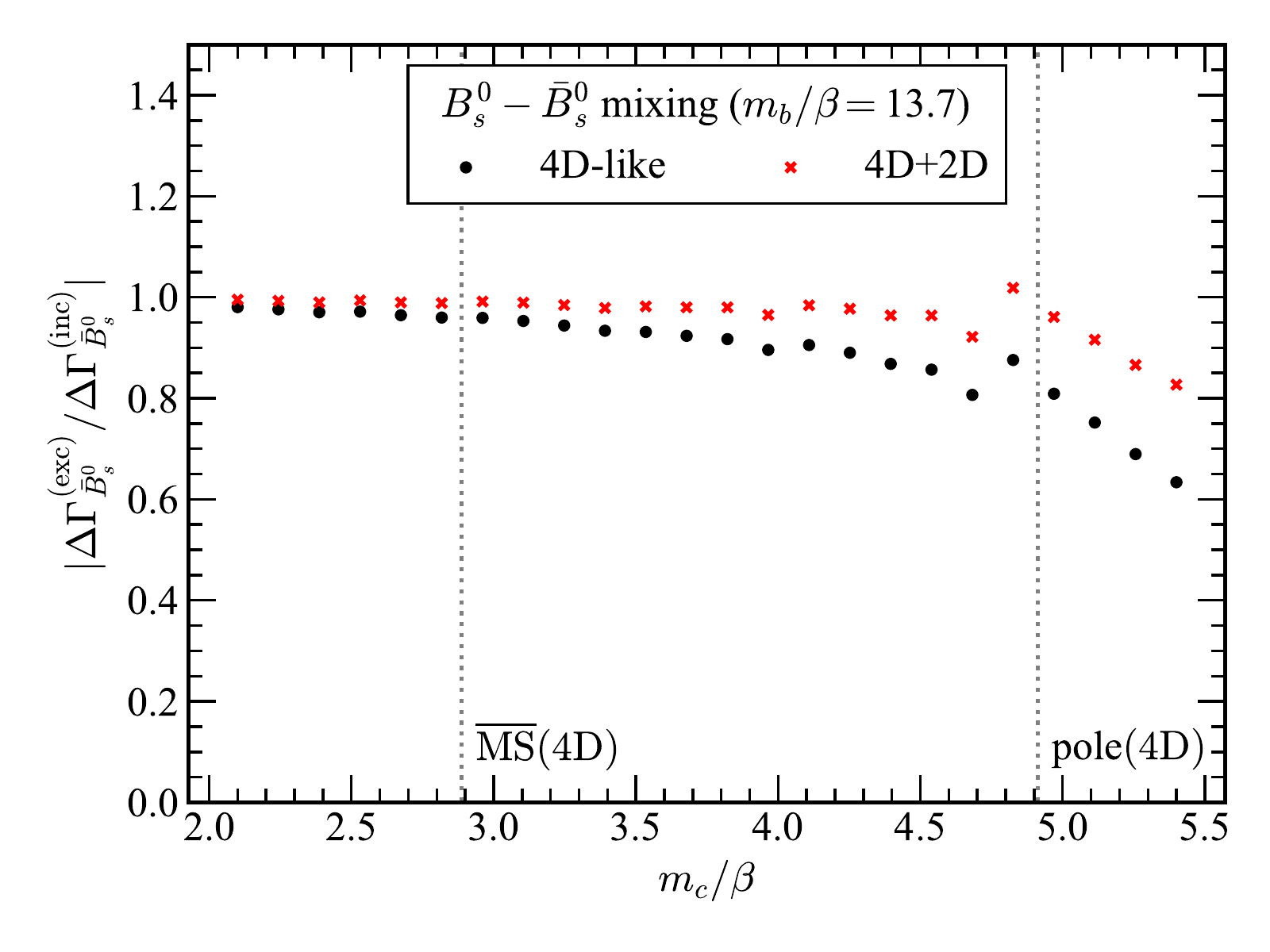}}\\
\subfigure[]{\includegraphics[width=0.95\columnwidth]{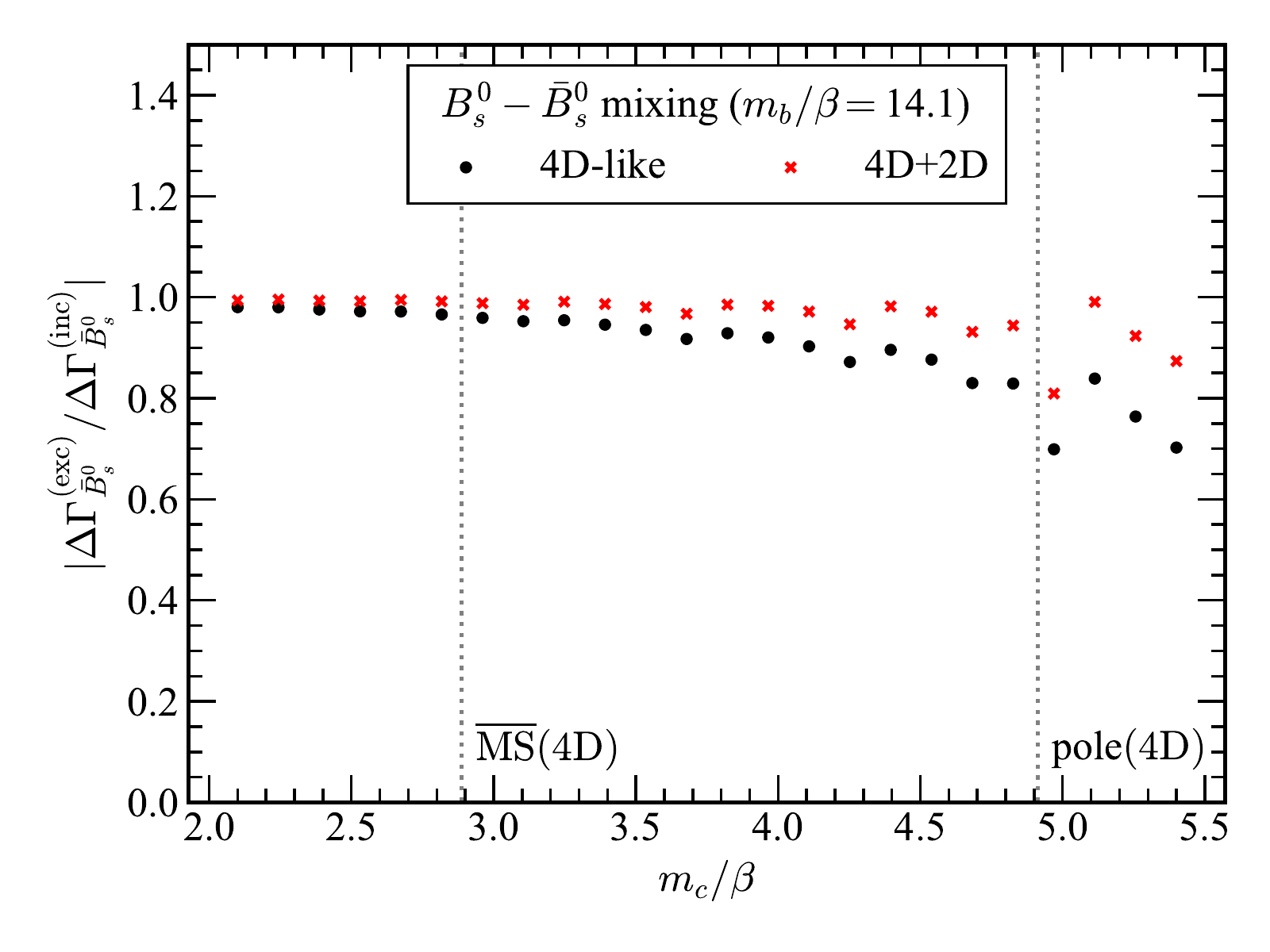}}
\end{center}
\vspace{-5mm}
\caption{Same as Fig.~\ref{Fig:9} for the $B^0_s-\bar{B}^0_s$ mixing: (a) $m_b/\beta=13.7$, (b) $m_b/\beta=14.1$.}
\label{Fig:11}
\end{figure}
\begin{figure}[H]
\vspace{-22mm}
\begin{center}
\subfigure[]{\includegraphics[width=0.95\columnwidth]{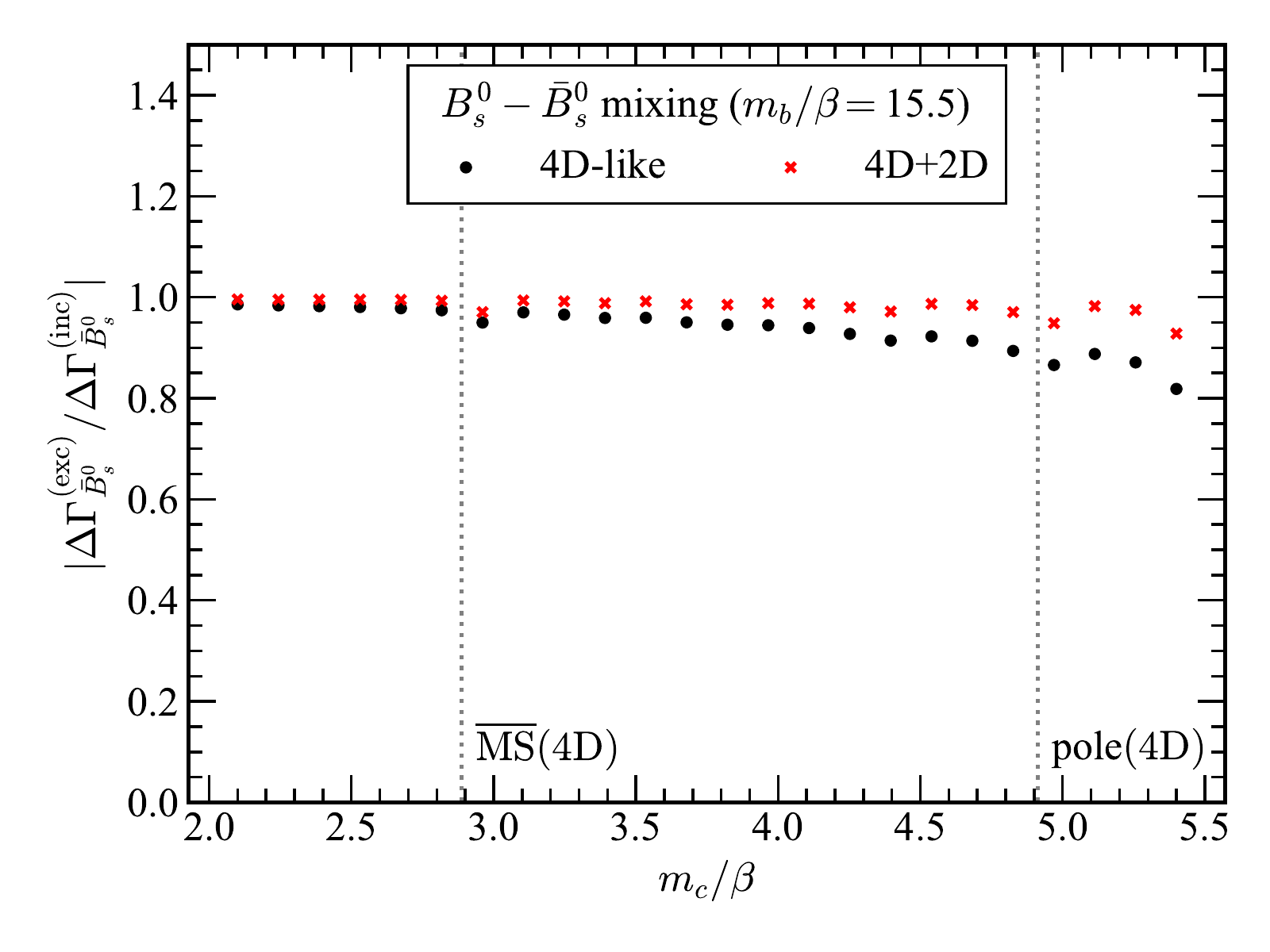}}\\
\subfigure[]{\includegraphics[width=0.95\columnwidth]{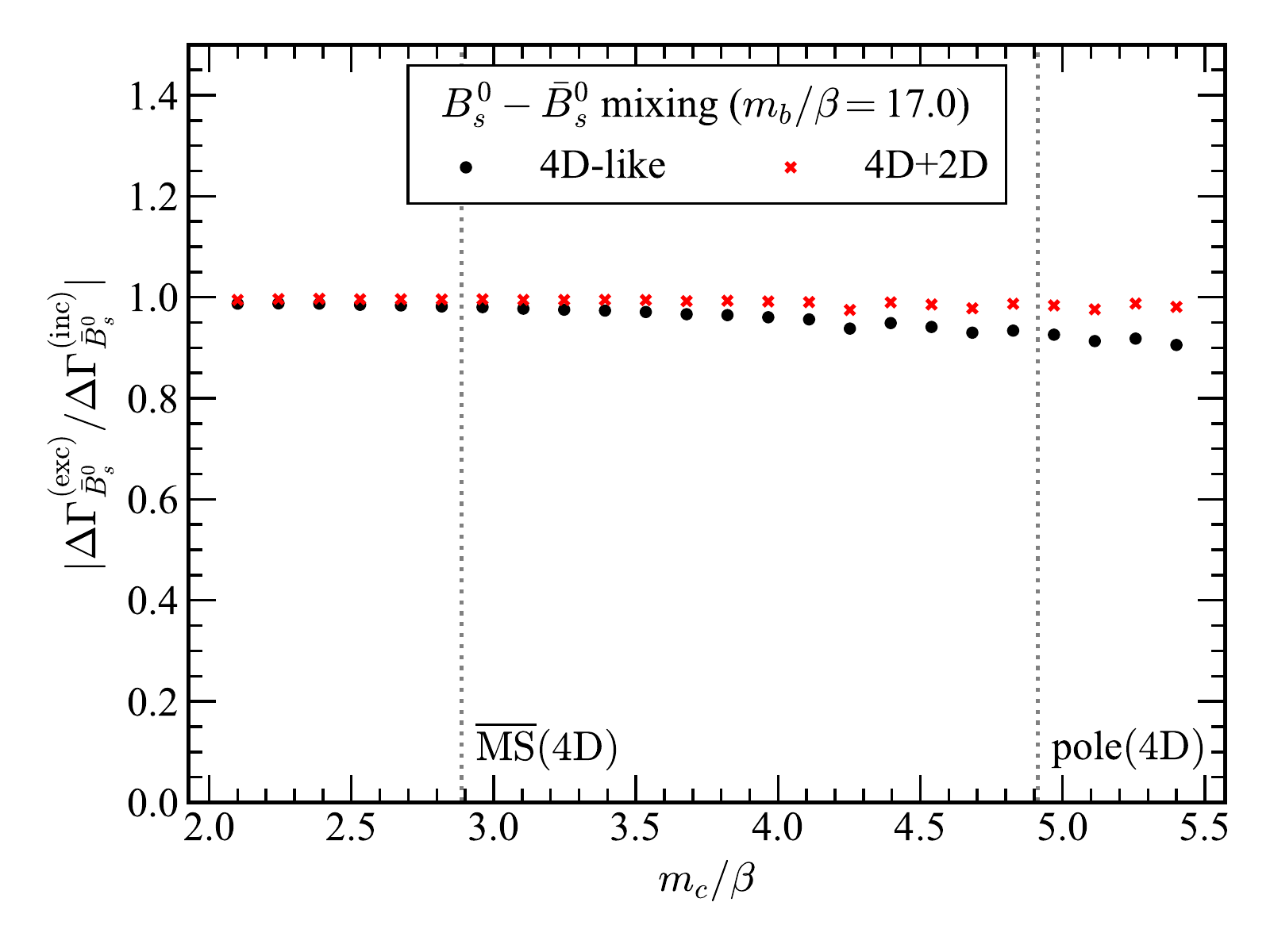}}
\end{center}
\vspace{-5mm}
\caption{Same as Fig.~\ref{Fig:9} for the $B^0_s-\bar{B}^0_s$ mixing: (a) $m_b/\beta=15.5$, (b) $m_b/\beta=17.0$.}
\label{Fig:12}
\end{figure}

The ratio of the inclusive observable to the sum of exclusive ones defined in Eqs.~(\ref{Eq:DelMG}, \ref{Eq:DelMGforB}) is shown in Figs.~\ref{Fig:9}-\ref{Fig:12} for the three meson mixings. In obtaining the figures, we included all the three terms in Eqs.~(\ref{Eq:G12incD}, \ref{Eq:G12incB}). The numerical results are stabilized as the second terms give quite small contributions. One finds that for the $D^0-\bar{D}^0$ mixing, the patterns in Fig.~\ref{Fig:9} are precisely similar to those in Fig.~\ref{Fig:6}, which are regarded as the cases in the limit of $\lambda_b\to 0$ in Fig.~\ref{Fig:9}. Hence, the net observable for the $D^0-\bar{D}^0$ mixing is enhanced when the phase space is given by one in four-dimensions, as well as Fig.~\ref{Fig:6}. Meanwhile, the patterns in Figs.~\ref{Fig:10}-\ref{Fig:11} for the $B_q^0-\bar{B}_q^0$ mixing are distinguished from those in Figs.~\ref{Fig:7}-\ref{Fig:8}: the enhancement in the order of magnitude does not occur in Figs.~\ref{Fig:10}-\ref{Fig:11}, yet the visible difference between inclusive and exclusive results exists. This gloss pattern is consistent with the realistic observation in the $D^0-\bar{D}^0$ and $B^0_q-\bar{B}^0_q (q=d, s)$ mixings. That the huge enhancement occurs solely for the $D^0-\bar{D}^0$ mixing is interpreted as the strong sensitivity to (GIM 1), unlike the $B^0_q-\bar{B}^0_q$ mixing, as seen in the approximate relations in Eqs.~(\ref{Eq:Appro1}, \ref{Eq:Appro2}).
\par
For the $B^0_q-\bar{B}^0_q$ mixing, further comparison between the four-dimensional observation and two-dimensional results is given in order. For $q=d$, the HFLAV result for $\Delta \Gamma_{B_d}$ \cite{Amhis:2019ckw} is consistent with zero within an error while the four-dimensional HQE result is given by $\Delta \Gamma_{B_d} = (2.6 \pm 0.4) \times 10^{-3}~\mathrm{ps}^{-1}$ \cite{Lenz:2019lvd}. Due to this situation in four-dimensions, a visible size of the correction to the HQE prediction in the $B_d^0-\bar{B}_d^0$ mixing is possible, being still consistent with the two-dimensional result in Fig~\ref{Fig:10}. As for $q=s$, by combining the results of the HFLAV \cite{Amhis:2019ckw} and the HQE \cite{Lenz:2019lvd}, one obtains a ratio, $\Delta \Gamma_{B_s}^{\rm (ex)}/\Delta \Gamma_{B_s}^{\rm (th)}=0.99\pm 0.15$ in four-dimensions (the error largely comes from the theoretical side). For the two-dimensional result, the correction to $|\Delta \Gamma_{B_s}^{\rm (exc)}/\Delta \Gamma_{B_s}^{\rm (inc)}|$ from unity is less than $20 \% ~ (18\%)$ for $m_b/\beta=13.7~(14.1)$ for the plotted points with $m_c <m_c^{\rm pole, 4D}$ in Fig.~\ref{Fig:11}. For this region of charm quark mass, the result in two-dimensions is consistent with what is currently indicated in four-dimensions. In order to check the region for larger bottom quark mass, the width differences with $m_b/\beta=15.5$ and $17.0$ are shown in Fig.~\ref{Fig:12} for the $B_s^0-\bar{B}^0_s$ mixing. One can find that the correction to $|\Delta \Gamma_{B_s}^{\rm (exc)}/\Delta \Gamma_{B_s}^{\rm (inc)}|$ from unity is less than $11\%$ $(8\%)$ for $m_b/\beta=15.5$ $(17.0)$ in the region of $m_c <m_c^{\rm pole, 4D}$, being consistent with the observation in four-dimensions within $1\sigma$.

\section{Conclusion}\label{Sec:5}\noindent
We have studied local quark-hadron duality and its violation in the heavy quark mixings on the basis of one certain dynamical mechanism. For the inclusive analysis, we have obtained the leading HQE expression that arises from the four-quark operators by evaluating the box diagrams in two-dimensions. The resulting width difference scales like a constant for large $m_Q$, with the correction to this starting from $1/m_Q$, which was clarified in the static limit. Care must be taken for the fact that, in the presence of the GIM mechanism, the order of magnitude for the inclusive observables strongly depends on whether the 4D-like phase space is solely considered or 2D-specific ones are also included.
\par
We have analytically shown that local duality is unambiguously seen in the massless limits for $u$  and $d$ quarks, which might be relevant for $D^0\to \pi^+\pi^-\to \bar{D}^0$ and $\bar{B}^0_d\to \pi^-\pi^+\to B^0_d$, by comparing the inclusive and exclusive width differences. This is interpreted as an example of the ``exclusive'' duality. For the massive case, duality violation is numerically investigated for the three meson mixings with the 't Hooft equation being solved. For two massive intermediate contributions, {\it e.g.}, $c\bar{u}\to s\bar{s}\to u\bar{c}$, the spikes for the exclusive width differences appear when the heavy quark mass gets larger than values at each kinematical threshold.  
\par
As stressed in the Introduction, the realistic observation in four-dimension indicates that the discrepancy between theory and experiment is of four orders of magnitude for the observable in the $D^0-\bar{D}^0$ mixing when the HQE result is given by the four-quark operators. In an attempt to interpret this observation, we have investigated how the exclusive observable is enhanced, relative to one obtained by the inclusive analysis, in the presence of the GIM mechanism. For the $D^0-\bar{D}^0$ mixing, the enhancement for the exclusive result is shown, confirmed to be larger than $10^3$ for $0.14<m_s/\beta<0.25$, when the phase space function is given by only the 4D-like term, although a huge enhancement is absent when the contributions of the 2D-specific phase space terms are added. As for the $B^0_q-\bar{B}^0_q$ $(q=d, s)$ mixing, no huge enhancement of the exclusive observable is realized, yet
the visible correction to $|\Delta \Gamma_{B_q}^{\rm (exc)}/\Delta \Gamma_{B_q}^{\rm (inc)}|$ from unity is seen, particularly arising from the $b\bar{q}\to c\bar{c}\to q\bar{b}$. Further improvement in the precision of the exclusive analysis is a technical task. If the domain of $m_c<m_c^{\rm pole, 4D}$ is considered, the correction to the ratio for the $B_s^0-\bar{B}_s^0$ mixing is typically less than $(20\%, 18\%, 11\%, 8\%)$ for $m_b/\beta=(13.7, 14.1, 15.5, 17.0)$, being still consistent with what is currently indicated in four-dimensions. Those non-negligible corrections to the HQE based on the most color-allowed topology motivate the future measurement in the $B_d^0-\bar{B}_d^0$, and suggest that the HQE prediction for $B_s^0-\bar{B}_s^0$ should be made more precise, in order to check whether non-negligible duality violation is seen.
\begin{acknowledgments}\noindent
The author would like to thank Hai-Yang Cheng, Hsiang-nan Li and Takuya Morozumi for reading the manuscript and useful comments. Part of the numerical computation in this project was performed by the computational resources at Academia Sinica Grid Computing Centre (ASGC). This work was supported in part by MOST of R.O.C. under Grant No.~MOST-107-2119-M-001-035-MY3.
\end{acknowledgments}
\appendix
\section{Box diagram in $1+1$ dimensions}
For $Q(p_1)\bar{q}(p_2)\to q(p_3)\bar{Q}(p_4)$, one finds that Fig.~\ref{Fig:1}a with internal quarks being labeled as $(i, j) =(d, d), (s, d), (s, s)$ for $c\bar{u}\to u\bar{c}$ and $(i, j) =(u, u), (c, u), (u, u)$ for $b\bar{q}\to q\bar{b}$ $(q=d, s)$ is calculated in $d$ dimensions,
\bea
\left.\mathcal{A}_{ij}\right|_{(\mathrm{a})}&=&-\lambda_i\lambda_j\left(\frac{-ig_2}{\sqrt{2}}\right)^4\int \frac{\D^dq}{(2\pi)^d i}\bar{q}(p_3)\gamma^\mu (c_\mathrm{V}+c_\mathrm{A}\gamma_5)\frac{1}{\slashed{q}-m_i+i\epsilon}\gamma^{\nu}(c_\mathrm{V}+c_\mathrm{A}\gamma_5)Q(p_1)\nn\\
&\times &\bar{q}(p_2)\gamma_\nu (c_\mathrm{V}+c_\mathrm{A}\gamma_5)\frac{1}{\slashed{q}-\slashed{p}_1-\slashed{p}_2-m_j+i\epsilon}\gamma_{\mu}(c_\mathrm{V}+c_\mathrm{A}\gamma_5)Q(p_4)\nn\\
&\times&\frac{1}{(q-p_1)^2-M_W^2+i\epsilon}\frac{1}{(q-p_3)^2-M_W^2+i\epsilon}\label{Eq:amp1},\qquad\qquad
\eea
The above expression is readily evaluated in the approximation where the momenta of heavy quark ($Q$) is much larger than ones of the spectator quark ($q$), {\it i.e.}, $p_1\gg p_2, p_4 \gg  p_3$. This is done by decomposing the product of the propagators into partial fractions \cite{Cheng:1982hq} with $x_\beta = m_\beta^2/M_W^2~(\beta=Q, i, j)$,
\bea
&&\frac{1}{q^2-m_i^2}\frac{1}{(q-p_1)^2-m_j^2}\frac{1}{(q-p_1)^2-M_W^2}\frac{1}{q^2-M_W^2}\nn\\
&&=\frac{1}{M_W^4(1-x_i^2)(1-x_j^2)}\left[\frac{1}{(q^2-m_i^2)[(q-p_1)^2-m_j^2]}+\frac{1}{(q^2-M_W^2)[(q-p_1)^2-M_W^2]}\right.\nn\\
&&\left.\qquad\qquad\qquad\qquad\qquad-\frac{1}{(q^2-m_i^2)[(q-p_1)^2-M_W^2]}-\frac{1}{(q^2-M_W^2)[(q-p_1)^2-m_j^2]}\right].\qquad\qquad \label{Eq:pafra}
\eea
Hereafter we suppress $1/[(1-x_i)(1-x_j)]$ in Eq.~(\ref{Eq:pafra}), which approaches unity in $M_W\to \infty$. By defining an object analogous to the Fermi constant, $G_F/\sqrt{2}=g_2^2/8M_W^2$, one gets,
\bea
\left.\mathcal{A}_{ij}\right|_{(\mathrm{a})}&=&-8\lambda_i\lambda_jG_F^2\left(\displaystyle\sum_{k=1}^{2}-\displaystyle\sum_{k=3}^{4}\right)\left\{\left[g_{\rho\sigma}F^{(k)}_{ij}-p_{1\rho} p_{1\sigma} G^{(k)}_{ij}\right]\right.\nn\\
&\times&[\bar{q}(p_3)\gamma_\mu \gamma^\rho\gamma_{\nu}(c_\mathrm{V}+c_\mathrm{A}\gamma_5)^2Q(p_1)][\bar{q}(p_2)\gamma^\nu \gamma^\sigma\gamma^{\mu}(c_\mathrm{V}+c_\mathrm{A}\gamma_5)^2Q(p_4)]\nn\\
&&+(c_\mathrm{V}^2-c_\mathrm{A}^2)^2m_im_jH^{(k)}_{ij}[\bar{q}(p_3)\gamma_\mu \gamma_{\nu}Q(p_1)][\bar{q}(p_2)\gamma^\nu \gamma^{\mu}Q(p_4)]\nn\\
&&-(c_\mathrm{V}^2-c_\mathrm{A}^2)m_ip_{1\rho}I^{(k)}_{ij}[\bar{q}(p_3)\gamma_\mu \gamma_{\nu}Q(p_1)][\bar{q}(p_2)\gamma^\nu \gamma^\rho\gamma^{\mu}(c_\mathrm{V}+c_\mathrm{A}\gamma_5)^2Q(p_4)]\nn\\
&&\left.+(c_\mathrm{V}^2-c_\mathrm{A}^2)m_jp_{1}^{\rho} I^{(k)}_{ji}[\bar{q}(p_3)\gamma_\mu \gamma_\rho\gamma_{\nu}(c_\mathrm{V}+c_\mathrm{A}\gamma_5)^2Q(p_1)][\bar{q}(p_2)\gamma^\nu \gamma^{\mu}Q(p_4)]\right\},
\qquad\quad\label{Eq:ampb}
\eea
where $F^{(k)}_{ij}, G^{(k)}_{ij}, H^{(k)}_{ij}$ and $I^{(k)}_{ij}$ are loop integrals given by,
\bea
F^{(k)}_{ij}&=&\int_0^1\D\alpha\int\frac{\D ^dq}{(2\pi)^d i}\frac{q^2/d}{[q^2-M_W^2\Lambda^{(k)}_{ij}(\alpha)]^2}=-\frac{1}{2}\frac{\Gamma(1-\frac{d}{2})}{(4\pi)^{d/2}}\int_0^1\D\alpha\left(\frac{1}{M_W^2\Lambda^{(k)}_{ij}(\alpha)}\right)^{1-d/2},\label{Eq:funF}\qquad\quad\\
G^{(k)}_{ij}&=&\int_0^1\D\alpha\int\frac{\D ^dq}{(2\pi)^d i}\frac{\alpha(1-\alpha)}{[q^2-M_W^2\Lambda_{ij}^{(k)}(\alpha)]^2}=
\frac{\Gamma(2-\frac{d}{2})}{(4\pi)^{d/2}}\int_0^1\D\alpha\:\alpha(1-\alpha)\left(\frac{1}{M_W^2\Lambda^{(k)}_{ij}(\alpha)}\right)^{2-d/2},\quad\qquad\\
H^{(k)}_{ij}&=&\int_0^1\D\alpha\int\frac{\D ^dq}{(2\pi)^d i}\frac{1}{[q^2-M_W^2\Lambda^{(k)}_{ij}(\alpha)]^2}=
\frac{\Gamma(2-\frac{d}{2})}{(4\pi)^{d/2}}\int_0^1\D\alpha\left(\frac{1}{M_W^2\Lambda^{(k)}_{ij}(\alpha)}\right)^{2-d/2},\quad\qquad\\
I^{(k)}_{ij}&=&\int_0^1\D\alpha\int\frac{\D ^dq}{(2\pi)^d i}\frac{\alpha}{[q^2-M_W^2\Lambda^{(k)}_{ij}(\alpha)]^2}=\frac{\Gamma(2-\frac{d}{2})}{(4\pi)^{d/2}}\int_0^1\D\alpha~\alpha\left(\frac{1}{M_W^2\Lambda^{(k)}_{ij}(\alpha)}\right)^{2-d/2},\quad\qquad \label{Eq:funI}
\eea
where some objects analogous to those in $3+1$ dimensions \cite{Cheng:1982hq, Buras:1984pq} are introduced,
\bea
\Lambda^{(1)}_{ij}&=&(1-\alpha) x_i+\alpha x_j-\alpha(1-\alpha)x_Q-i\epsilon,\\
\Lambda^{(2)}_{ij}&=&1-\alpha(1-\alpha)x_Q-i\epsilon,\\
\Lambda^{(3)}_{ij}&=&(1-\alpha) x_i+\alpha -\alpha(1-\alpha)x_Q-i\epsilon,\\
\Lambda^{(4)}_{ij}&=&(1-\alpha) +\alpha x_j-\alpha(1-\alpha)x_Q-i\epsilon.
\eea
There are a few points to be mentioned. First, due to the asymmetric sum of $k$, terms independent of $k$ vanish in Eq.~(\ref{Eq:ampb}). Second, a threshold relevant for two internal quarks is associated with $\Lambda_{ij}^{(1)}$ in Eq.~(\ref{Eq:ampb}) while $\Lambda^{(3)}_{ij}$ and $\Lambda^{(4)}_{ij}$ ($\Lambda^{(2)}_{ij}$) correspond to that of the single (double) $W$ boson(s). Thus, only $k=1$ in Eq.~(\ref{Eq:ampb}) is of our current interest to calculate the absorptive part. Third, for $d=2$, all of the functions in Eqs.~(\ref{Eq:funF}-\ref{Eq:funI}) give rises to discontinuity, contributing to the width difference. As we will see later, the discontinuities of $G_{ij}^{(k)}, H_{ij}^{(k)}$ and $I_{ij}^{(k)}$ have function forms distinct from ones for $d=4$.\par
Assembling the above-mentioned points, and fixing $d=2$, we take the finite contributions in Eq.~(\ref{Eq:ampb}),
\bea
\left.\mathcal{A}_{ij}\right|_{(\mathrm{a})}&=&-\lambda_i\lambda_j\frac{G_F^2}{\pi}\left\{\left[g_{\rho\sigma}\bar{F}_{ij}-2\frac{p_{1\rho}p_{1\sigma}}{m_Q^2}\bar{G}_{ij}\right]\right.\nn\\
&&\times[\bar{q}(p_3)\gamma_\mu \gamma^\rho\gamma_{\nu}(c_\mathrm{V}+c_\mathrm{A}\gamma_5)^2Q(p_1)] [\bar{q}(p_2)\gamma^\nu \gamma^\sigma\gamma^{\mu}(c_\mathrm{V}+c_\mathrm{A}\gamma_5)^2Q(p_4)]\nn\\
&&+2(c_\mathrm{V}^2-c_\mathrm{A}^2)^2\bar{H}_{ij}[\bar{q}(p_3)\gamma_\mu \gamma_{\nu}Q(p_1)][\bar{q}(p_2)\gamma^\nu \gamma^{\mu}Q(p_4)]\nn\\
&&-2(c_\mathrm{V}^2-c_\mathrm{A}^2)\frac{p_{1\rho}}{m_Q} \bar{I}_{ij}[\bar{q}(p_3)\gamma_\mu \gamma_{\nu}Q(p_1)][\bar{q}(p_2)\gamma^\nu \gamma^\rho\gamma^{\mu}(c_\mathrm{V}+c_\mathrm{A}\gamma_5)^2Q(p_4)]\nn\\
&&\left.+2(c_\mathrm{V}^2-c_\mathrm{A}^2)\frac{p_{1}^{\rho}}{m_Q} \bar{I}_{ji}[\bar{q}(p_3)\gamma_\mu \gamma_\rho\gamma_{\nu}(c_\mathrm{V}+c_\mathrm{A}\gamma_5)^2Q(p_1)][\bar{q}(p_2)\gamma^\nu \gamma^{\mu}Q(p_4)]
\right\},\qquad\quad\label{Eq:amp2}
\eea
where the functions that have branch cut are introduced by,
\bea
&\bar{F}_{ij}=\displaystyle\int_0^1\ln(M_W^2 \Lambda^{(1)}_{ij})\D\alpha,\quad
\bar{G}_{ij}=m_Q^2\int_0^1\frac{\alpha(1-\alpha)\D\alpha }{M_W^2 \Lambda^{(1)}_{ij}},&\nn\\
&\bar{H}_{ij}=m_im_j\displaystyle\int_0^1 \frac{\D\alpha}{M_W^2 \Lambda^{(1)}_{ij}},\quad
\bar{I}_{ij}=m_im_Q\displaystyle\int_0^1 \frac{\alpha \D\alpha}{M_W^2 \Lambda^{(1)}_{ij}}.\quad
&
\label{Eq:functions}
\eea
The discontinuities of Eq.~(\ref{Eq:functions}) in a physical region are (the sign is associated with ones above branch cut),
\bea
&\mathrm{Disc}\:\bar{F}_{ij}=-2\pi iF^{\rm (th)}_{ij},\quad 
\mathrm{Disc}\:\bar{G}_{ij}=+2\pi i G^{\rm (th)}_{ij},&\nn\\
&\mathrm{Disc}\:\bar{H}_{ij}=+4\pi i H^{\rm (th)}_{ij},\quad
\mathrm{Disc}\:\bar{I}_{ij}=+2\pi i I^{\rm (th)}_{ij}.\quad 
&
\label{Eq:Disc1}
\eea
where $F^{\rm (th)}_{ij}, G^{\rm (th)}_{ij}, H^{\rm (th)}_{ij}$ and $I^{\rm (th)}_{ij}$ are defined in Eqs.~(\ref{Eq:phasespace0}-\ref{Eq:phasespace}).
\par
The terms proportional to $\bar{F}_{ij}$ and $\bar{H}_{ij}$ in Eq.~(\ref{Eq:amp2}) are facilitated by the Fiertz rearrangement in two-dimensions,
\bea
[\overline{\psi}_1 \gamma^\mu \gamma^\rho\gamma^\nu (c_\mathrm{V}+c_\mathrm{A}\gamma_5)^2\psi_2][\overline{\psi}_3 \gamma_\nu \gamma_\rho\gamma_\mu (c_\mathrm{V}+c_\mathrm{A}\gamma_5)^2\psi_4]&\stackrel{2D}{=}&-4(c_\mathrm{V}^2-c_\mathrm{A}^2)^2(\overline{\psi}_1 \gamma^\mu \gamma_5\psi_2)(\overline{\psi}_3 \gamma_\mu  \gamma_5\psi_4),\label{Eq:Fiertz1}\quad\qquad\;\;
\eea
\vspace{-9mm}
\bea
[\overline{\psi}_1 \gamma^\mu \gamma^\nu \psi_2][\overline{\psi}_3 \gamma_\nu \gamma_\mu \psi_4]&\stackrel{2D}{=}&2[(\overline{\psi}_1  \psi_2)(\overline{\psi}_3 \psi_4)-(\overline{\psi}_1  i\gamma_5\psi_2)(\overline{\psi}_3i\gamma_5 \psi_4)],\label{Eq:Fiertz2}
\eea
where in Eq.~(\ref{Eq:Fiertz1}) we used $\gamma_\mu=\epsilon_{\mu\nu}\gamma^\nu\gamma_5~(\epsilon_{01}=+1)$ valid in two-dimensions, that yields $V^\mu\times V_\mu= -A^\mu\times A_\mu$. As for the terms proportional to $\bar{G}_{ij}$ and $\bar{I}_{ij}$, the relevant Fiertz rearrangements are also obtainable straightforwardly, with the equation of motion for heavy quark being implemented. Below, we omit the bilinears that do not contribute to heavy meson mixings for the ground state in the large-$N_c$ limit. By substituting Eqs.~(\ref{Eq:Disc1}-\ref{Eq:Fiertz2}) into Eq.~(\ref{Eq:amp2}), we obtain the absorptive part of Fig.~\ref{Fig:1}a,
\bea
\left.\mathrm{Disc}\:\mathcal{A}_{ij}\right|_{(\mathrm{a})}&\to&-8i\lambda_i\lambda_jG_F^2(c_{\rm V}^2-c_{\rm A}^2)\left\{\left[(c_{\rm V}^2-c_{\rm A}^2)\left(F^{\rm (th)}_{ij}+2G^{\rm (th)}_{ij}\right)-(c_{\rm V}^2+c_{\rm A}^2)\left(I^{\rm (th)}_{ij}+I^{\rm (th)}_{ji}\right)\right]\right.\nn\\
&&\times\left. [\bar{q}(p_3)\gamma^\mu\gamma_5 Q(p_1)][\bar{q}(p_2)\gamma_\mu \gamma_5Q(p_4)]\right.\nn\\
&&-\left[(c_{\rm V}^2-c_{\rm A}^2)\left(G^{\rm (th)}_{ij}+2H^{\rm (th)}_{ij}\right)+(c_{\rm V}^2+c_{\rm A}^2)\left(I^{\rm (th)}_{ij}+I^{\rm (th)}_{ji}\right)
\right]\nn\\
&&\left.\times[\bar{q}(p_3) i\gamma_5Q(p_1)][\bar{q}(p_2)i\gamma_5 Q(p_4)]
\right\},\label{Eq:absa}\qquad\qquad
\eea
Thus, the contribution of the $V\pm A$ current, corresponding to $c_\mathrm{V}=\pm c_\mathrm{A}$, vanishes for $g_{\mu\nu}$ part of the $W$ propagator. This point is distinct from the familiar case in four-dimensions, where the Fiertz rearrangement gives,
\bea
[\overline{\psi}_1 \gamma^\mu \gamma^\rho\gamma^\nu (1\pm \gamma_5)\psi_2][\overline{\psi}_3 \gamma_\nu \gamma_\rho\gamma_\mu (1\pm \gamma_5)\psi_4]\stackrel{4D}{=}4[\overline{\psi}_1 \gamma^\mu (1\pm \gamma_5) \psi_2][\overline{\psi}_3 \gamma_\mu (1\pm \gamma_5) \psi_4],\qquad
\eea
so that (part of) the final result is proportional to the $(V\pm A)\times (V\pm A)$ operator in four-dimensions. This difference is due to the vanishing of $\gamma_\mu \gamma_\alpha \gamma^\mu=(2-d) \gamma_\alpha$, and also to the higher redundancy for products of gamma matrices for $d=2$ than that for $d=4$.
\par
Likewise, one can also calculate the absorptive part of Fig.~\ref{Fig:1}b, which gives the amplitude similar to Eq.~(\ref{Eq:absa}) except that the momentum arrangement for the spinors is different. By combining these results, we finally obtain the effective Hamiltonian in Eq.~(\ref{Eq:Hham}) with Eqs.~(\ref{Eq:G12def1}, \ref{Eq:BBbar}).

\end{document}